\ifdefined\Anonymous
\documentclass[sigconf,anonymous]{acmart}
\else
\documentclass[sigconf]{acmart}
\fi
\usepackage[noend]{algpseudocode}

\usepackage{algorithm}
\usepackage[normalem]{ulem}
\usepackage{natbib}
\usepackage{framed}
\usepackage{subfigure}
\def\D{\mathcal{D}}
\def\X{\mathcal{X}}
\def\Y{\mathcal{Y}}
\def\W{\mathcal{W}}
\def\L{\mathcal{L}}
\def\T{\mathcal{T}}
\def\B{\mathcal{B}}
\def\E{\mathcal{E}}
\def\G{\mathcal{G}}
\def\H{\mathcal{H}}
\def\I{\mathcal{I}}
\def\t{\mathrm{t}}

\def\eps{\epsilon}
\newtheorem{observation}{Observation}

\def\acc{\texttt{Accept}}
\def\rej{\texttt{Reject}}
\def\added#1{{#1}}
\AtBeginDocument{%
  }

\begin{document}
\setcopyright{none}
\title{Proof-of-Learning with Incentive Security}


\author{Zishuo Zhao}
\affiliation{%
  \institution{National University of Singapore}
  \country{Singapore}}
\email{wiku30@nus.edu.sg}

\author{Zhixuan Fang}
\affiliation{%
  \institution{Tsinghua University}
  \institution{Shanghai Qi Zhi Institute}
  \city{Beijing}
  \country{China}}
\email{zfang@mail.tsinghua.edu.cn}

\author{Xuechao Wang}
\affiliation{%
  \institution{Thrust of Fintech}
  \institution{HKUST (GZ)}
  \city{Guangzhou}
  \country{China}}
\email{xuechaowang@hkust-gz.edu.cn}

\author{Xi Chen}
\affiliation{%
  \institution{New York University}
  \city{New York City}
  \country{United States}}
\email{xc13@stern.nyu.edu}

\author{Hongxu Su}
\affiliation{%
  \institution{HKUST (GZ)}
  \city{Guangzhou}
  \country{China}}
\email{hsu238@connect.hkust-gz.edu.cn}

\author{Haibo Xiao}
\affiliation{%
  \institution{Tsinghua University}
  \city{Beijing}
  \country{China}}
\email{xiaohb21@mails.tsinghua.edu.cn}

\author{Yuan Zhou}
\affiliation{%
  \institution{Yau Mathematical Sciences Center}
  \institution{Tsinghua University}
  \city{Beijing}
  \country{China}}
\email{yuan-zhou@tsinghua.edu.cn}


\begin{abstract}
Most concurrent blockchain systems rely heavily on the Proof-of-Work (PoW) or Proof-of-Stake (PoS) mechanisms for decentralized consensus and security assurance. However, the substantial energy expenditure stemming from computationally intensive yet meaningless tasks has raised considerable sustainability concerns surrounding traditional PoW approaches, and the PoS mechanism, while free of energy consumption, is subject to security and economic issues. Addressing these issues, the paradigm of Proof-of-Useful-Work (PoUW) seeks to employ challenges of practical significance as PoW, thereby imbuing energy consumption with tangible value. On the other hand, the trustworthiness of the training processes is also crucial for the initiative of Decentralized AI (DeAI). While previous efforts in Proof of Learning (PoL) explored the utilization of deep learning model training {and verification} of Stochastic Gradient Descent (SGD) tasks as PoUW challenges, recent research has revealed its vulnerabilities to adversarial attacks and the theoretical hardness in crafting a byzantine-secure PoL mechanism.

In this paper, we introduce the concept of \emph{incentive-security} that incentivizes rational provers to behave honestly for their best interest, bypassing the existing hardness to design a PoL mechanism with computational efficiency, a provable incentive-security guarantee, and controllable difficulty.  
For a training task of $E$ epochs and model size $|\W|$, we improve the relative computational overhead from $\Theta(1)$ to $O(\frac{\log E}{E})$ without any staking requirement, or $O(\frac{1}{E})$ with a staking requirement comparable to the block reward, and improve the communication complexity from $\Theta(E|\W|)$ to $O(E+|\W|\log E)$ or $O(E+|\W|)$, respectively. Furthermore, while most recent research on PoUW assumes trusted problem providers and verifiers, our design also guarantees 
\emph{verifier incentive-security} that bypasses the Verifier's Dilemma via a \emph{capture-the-flag} protocol. By securing ML training with provable incentive-security guarantees, our research not only proposes an eco-friendly solution to blockchain systems, but also provides a proposal for a completely decentralized computing power market in the new AI age. 

\end{abstract}

\ifdefined\Anonymous
\else
\begin{CCSXML}
<ccs2012>
   <concept>
       <concept_id>10002978.10003006.10003013</concept_id>
       <concept_desc>Security and privacy~Distributed systems security</concept_desc>
       <concept_significance>500</concept_significance>
       </concept>
   <concept>
       <concept_id>10010147.10010257.10010293.10010294</concept_id>
       <concept_desc>Computing methodologies~Neural networks</concept_desc>
       <concept_significance>300</concept_significance>
       </concept>
   <concept>
       <concept_id>10002978.10003029.10003031</concept_id>
       <concept_desc>Security and privacy~Economics of security and privacy</concept_desc>
       <concept_significance>500</concept_significance>
       </concept>
 </ccs2012>
\end{CCSXML}
\ccsdesc[500]{Security and privacy~Economics of security and privacy}
\ccsdesc[300]{Security and privacy~Distributed systems security}
\ccsdesc[300]{Computing methodologies~Neural networks}
\fi

\keywords{blockchain, mechanism design, {decentralized AI}, secure computation, trustworthy machine learning, sustainability}



\maketitle

\iffalse

\added{[DONE: Discussion on model theft.]}

\added{[TODO: specify additional problems needed to solve before deploying (e.g., communication cost)]}

\added{[TODO: in-detail discussion on coverage of attacks]}

\added{[DONE: collusion proofness and dishonest proofness (simple peer prediction)]}

\added{[DONE: motivation: prevent BOTH model spoofing and cheating w/ less computation]}

\added{[DONE: reward rule---preventing blind guessing of flags (within pp.)]}

\fi

\section{Introduction}
\label{sec:intro}

Blockchain, with prevailing examples as Bitcoin \citep{nakamoto2008bitcoin} and Ethereum \citep{buterin2014next}, is an emerging technology that maintains decentralized consensus via a distributed ledger that utilizes cryptographic techniques to achieve trust and security. To prevent sybil attacks in the consensus mechanism, the earliest and most conventional way is Proof-of-Work (PoW) \citep{jakobsson1999proofs,10.1145/2976749.2978341,gervais2016security,kiayias2020non} as Bitcoin uses: all ``miners'' attempt to solve a hash puzzle and the first miner getting a valid solution wins access to the block.

However, the huge and inefficient use of energy and severe carbon footprint in the traditional PoW mechanism draws wide concern and is recognized as heavily controversial for the environmental impact of the blockchain system \citep{vranken2017sustainability,stoll2019carbon}. Since May 2021, cryptocurrency mining and even cryptocurrency trading have been banned in China due to the ecological concern of energy inefficiency \citep{riley2021current}. To address the energy issue, researchers propose alternative consensus mechanisms, e.g.~Proof-of-Stake (PoS) \citep{kiayias2017ouroboros,10.1007/978-3-030-17653-2_23,10.1093/rfs/hhaa075} in order to substitute PoW, but they tend to have inherent drawbacks in security and centralization issues \citep{bagaria2022proof}. In the high-level view of economics, \citet{piketty2014capital} argued that the phenomenon of $r>g$, i.e.~the return rate on capital (``stake'') being greater than the rate of economic growth (``work''), results in wealth concentration and social instability. Indeed, the heavy computation cost arguably binds the voting power with real-world productivity rather than intangible tokens. Were the computation made useful, the Proof-of-Useful-Work (PoUW) mechanism would indeed resolve the energy issue while preserving the decentralization and security of PoW {\cite{hoffmann2022challenges,10.1145/3524458.3547248}}. On the other hand, there are also positive views on the energy consumption of PoW mechanisms, e.g. the expansion of energy demand also motivates the development of new energy solutions \cite{ibanez2023bitcoin}. Since our PoUW mechanism essentially improves the efficiency of energy consumption instead of eliminating it, in contrast to PoS, our mechanism preserves this social benefit of PoW in the meantime of improving its sustainability.

In the age in which artificial intelligence (AI) has been becoming one of the most attractive topics in modern technology, researchers are actively attempting to incorporate machine learning tasks as PoUW challenges, i.e.~Proof-of-Learning (PoL). 
{
As a consensus mechanism for the blockchain system, an ideal design of PoUW should satisfy the following properties:

\begin{enumerate}
    \item \textbf{Security}: For the security and credibility of the blockchain system, an ideal PoUW mechanism should have theoretically provable security guarantees against dishonest behavior.
    \item \textbf{Efficiency}: An ideal PoUW mechanism should have a low computational overhead {(redundancy)} for energy efficiency, as a main motivation of PoUW.
    \item \textbf{Controllable Difficulty\footnote{The ``difficulty'' of a PoW challenge can be defined as the (expected) amount of computation needed to solve it.}}: As a stable block production time (BPT) is essential for the blockchain system's stability \citep{zheng2020difficulty}, an ideal PoUW mechanism should use challenges with predictable and controllable difficulty.
\end{enumerate}

However, although there have been a series of PoL proposals in the literature (e.g., \cite{jia2021proof,baldominos2019coinai,liu2021proof,bravo2019proof}), as far as we are concerned, none of them could simultaneously satisfy the three properties above.} Particularly, the methodologies of existing PoL mechanisms can be organized into two classes:

\begin{enumerate}
    \item \textbf{Proof-of-Computation}: Proving that the training task is honestly done, e.g.~\citep{jia2021proof};
    \item \textbf{Proof-of-Performance}: Proving that the output model satisfies required accuracy on a test dataset, e.g.~\citep{baldominos2019coinai, liu2021proof,bravo2019proof}. 
\end{enumerate}

The difficulty of designing a desirable PoL mechanism is observed as follows. For Proof-of-Computation mechanisms, a recent work \citep{fang2023proof} shows the hardness of efficiently verifying the correctness of a Proof-of-Computation with provable security guarantees without a further theoretical understanding of deep learning --- {particularly, the work of \citet{jia2021proof} is subject to adversarial attacks \citep{zhang2022adversarial,fang2023proof}.} For Proof-of-Performance mechanisms, \citet{hoffmann2022challenges} argues that it is hard to {evaluate} the actual difficulty (even possibility) to achieve given accuracy, leading to a barrier to controllable difficulty. In summary of the existing PoL mechanisms, we observe a Trilemma of Proof-of-Learning as below:

\begin{framed}
\begin{center}
%\begin{large}
\textbf{\emph{Trilemma of Proof-of-Learning}}
%\end{large}
\end{center}
    It is difficult to design a Proof-of-Learning mechanism that simultaneously satisfies \textbf{\underline{perfect security}}, \textbf{\underline{efficiency}} and \textbf{\underline{controllable difficulty}}.
\end{framed}

{ 
%Nevertheless, the verification of machine learning tasks is much trickier than the verification of hash puzzles, and it generally requires a party as the \emph{verifier} to verify the validity of the PoL. 

%Even with verifiers, recent works show a trilemma of Proof-of-Learning (PoL): it is difficult, if not impossible, to design a PoL challenge that simultaneously ensures satisfactory security, efficiency and controllable difficulty. 
}

In this research, we are motivated to resolve the sustainability issue of blockchain systems via a Proof-of-Computation mechanism to machine learning model training, and tackle the trilemma via a delicate relaxation of the security notion. Instead of preventing all attacks from being conducted without getting detected (byzantine security), we aim to prevent the attacks from ``being useful'' 
\ifdefined\enableAIS
{
with the following notions:

\begin{itemize}
    \item basic incentive-security: an attacker cannot increase their utility via saving computational cost by cheating.
    \item model-security: an attack attempting to submit a substantially incorrect model will be detected with high probability.
\end{itemize}
}
\else{}
with the \emph{incentive-security} notion, i.e.~an attacker cannot increase their utility via saving computational cost by cheating.
Particularly, our mechanism {in which the prover trains with designated random seeds and the verifier verifies random subsets of stages (as shown in Section~\ref{subsec:prover:prot}-\ref{subsec:verifier:prot})} can prevent the attacks of \cite{zhang2022adversarial} and \cite{fang2023proof} in the way as follows. From the stochastic nature of SGD, the verification protocol of \cite{jia2021proof} introduces a ``tolerance'' that allows small discrepancies in verification, which is exploited by these attacks. As our mechanism replaces the tolerance with designated random seeds\footnote{Different types of machines or softwares may have different rounding behavior, but we can enforce high precision and set a tolerance low enough to prevent any ``meaningful'' attack.}, our mechanism is enabled to catch their exploits as ``dishonest stages'' successfully. Furthermore, our verification mechanism only has an $O(\frac{\log E}{E})$ relative computational overhead\footnote{The ratio of computational power consumption in verification to computational power consumption in model training.} for a total of $E$ epochs with no staking requirement, or $O(\frac{1}{E})$ with a staking requirement comparable to the block reward, compared to $\Theta(1)$ in the work of \citet{jia2021proof}. For a model of size $|\W|$, we also improves the communication complexity from $\Theta(E|\W|)$ to $O(E+|\W|\log E)$ or $O(E+|\W|)$, respectively.

From another perspective, {the recently rapid development of AI technologies also draws safety concerns on the trustworthiness of AI models \citep{bengio2024managing,bengio2024international,hendrycks2024introduction,ren2024safetywashing}. While studies on AI \emph{alignment} (e.g., \citep{ji2023ai,firt2023calibrating,lyu2024keeping}) address the internal risks of \emph{unrobust} AI models, attacks by malicious trainers via corrupting the training process may bypass the alignment measures. As a recent example, the adversarial attack on ByteDance LLM training by an intern, which leads to a huge loss \cite{BytedanceAttack}, draws attention to the systematic security of AI model training.} Compared to the Proof-of-Performance paradigm, our Proof-of-Computation mechanism offers additional practical value {as a decentralized surveillance measure of AI model training}. While the Proof-of-Performance mechanism is primarily motivated by the goal of improving the sustainability of blockchain PoW mining, thus \emph{\textbf{improving blockchain with AI}}, the Proof-of-Computation mechanism can also serve as a blockchain-based trustworthy AI platform, enhancing the security and credibility of machine learning, i.e. simultaneously \emph{\textbf{securing AI with blockchain}}.

Furthermore, while most recent research papers on PoUW explicitly or implicitly assume that problem providers are trusted --- so that their proposed system is not completely decentralized, we are also motivated to consider \emph{frontend incentive-security} against known-model and model-stealing attacks even when problem providers and provers are both untrusted, thus enabling full decentralization and more robustness of the system. (See discussion in Section~\ref{subsec:untrusted:problems})

Since the computational overhead of verification is low, our PoL protocol can be used for general applications in which the task provider would like to delegate the training/fine-tuning tasks for remote computation, as a Machine-Learning-as-a-Service (MLaaS) platform. Nevertheless, the functionality of verification makes the protocol particularly suitable for applications in which credibility of the model and/or training process is critical. Examples include AI grading \cite{li2023wrong,tomic2022ai}, where the transparency and accuracy of the grading model are essential for educational and hiring processes, and credit evaluation \cite{bhatore2020machine,langenbucher2022ai}, where the fairness and reliability of the model impact financial decisions. These applications benefit from PoL’s verification mechanism, ensuring that the models are trained correctly and securely, thereby enhancing trust in their outputs.

%In other words, with the Incentive Security guarantee, a rational prover would prefer to compute the task honestly to maximize their utility; if the Model Security is additionally satisfied, then the correctness of the model can be guaranteed even in the presence of (irrational) malicious provers.

In light of the security desiderata discussed above, in our paper, we propose an incentive-secure Proof-of-Learning mechanism with the following contributions  
\ifdefined\enableAIS{is three-fold}\else{consisting of}:
\begin{enumerate}
\item With trusted verifiers (that are widely assumed in previous works), we propose  our \emph{interactive-proof}-based basic design satisfying computational efficiency, controllable difficulty, and incentive-security against dishonest provers for any stochastic optimization tasks, e.g.~stochastic gradient descent (SGD), and also substantially improves the relative computational overhead of the previous work \cite{jia2021proof}. (Sections~\ref{sec:prelim}-\ref{section:overview}) \added{We additionally address the \emph{model theft} issue that appear in existing PoL designs. (Section~\ref{sss:theft}).} 
\ifdefined\enableAIS{
\item For SGD tasks that satisfy convexity and second-order Lipschitz conditions, we propose an augmentation to guarantee the model-security of our mechanism.
}\else{}
\item With untrusted verifiers, we propose a \emph{capture-the-flag} (CTF)
%\footnote{The term ``Capture-The-Flag (CTF)'' refers to a cybersecurity competition in which participants attempt to find ``flags'' hidden in deliberately vulnerable systems to claim rewards, which is conceptually similar to our design.} 
protocol that preserves all desired properties above in our basic design and additionally achieves incentive-security against dishonest verifiers and collusion-proofness against colluding participants. (Section~\ref{sec:verIS})
\item We prove the theoretical incentive-security properties of our mechanisms. (Section~\ref{sec:basicIS})
\end{enumerate}

Then, in Section~\ref{sec:exp}, we perform experimental evaluations to show the performance of our mechanism on real-world ML tasks. 

In Appendix~\ref{app:anomaly}, we further discuss on potential augmentations of our mechanism to ensure model correctness against malicious attacks even from irrational attackers.

%\textbf{Rounds of interaction.} Our basic mechanism needs one round of interaction between the prover and the verifier, and the full mechanism needs two rounds of interaction.

%\textbf{Limitation of our incentive model.} \added{[REVISE]} While our novel modeling of \emph{incentive-security} is a suitable relaxation both due to the Trilemma of Proof-of-Learning and
%the nature of blockchain systems whose security depends on economic incentives, our study focuses on the model of \emph{individually} rational parties and does not consider collusions between the prover and the verifier. Nevertheless, the anonymity of blockchain reduces the risk of collusion due to the difficulty for the prover to predict or identify the identity of the verifier, and we would leave the expansion of more general incentive models {with collusion-proofness} for future study.

%\vspace{-0.7em}
\section{Background and Related Work}

\subsection{PoUW and PoL in Literature}

The biggest concern of the traditional PoW mechanism is the computation, and essentially, energy consumption. %As discussed by \cite{gervais2016security}, the security of a PoW network, for example, against double-spending, relies on malicious parties controlling less than $50\%$ of total computational power on the network, i.e. high honest computational power is required for the system security; on the other hand, the high price of Bitcoin also attract miners consuming large amounts of energy in the mining competition. 
As discussed by \citet{chen2022crowdsourcing}, the current energy consumption of the Bitcoin network is around 120TWh per year, comparable to a medium-sized country, but the consumption serves no social welfare apart from maintaining the security scheme, leading to severe social inefficiency. In recent years, the wasteful energy consumption of blockchains, particularly Bitcoin, has been widely criticized around the world. Particularly, \citet{vranken2017sustainability} empirically discovered that the energy consumption of Bitcoin market is higher than its long-term benefit; \citet{stoll2019carbon} also noticed the severe carbon footprint of Bitcoin for sustainability issues. 

Aware of the energy and sustainability issues, previous research studied a wide variety of real-world problems that may serve as Proof-of-Useful-Work (PoUW) challenges. \citet{hoffmann2022challenges} surveyed the existing projects that incorporate number-theoretical, biological and machine learning problems into the PoUW mechanism. The survey shows a ``more usefulness, more challenge'' phenomenon in the existing works: while the Primecoin \citep{king2013primecoin} has been the most developed and already deployed on chain, the number-theoretical problem may be of limited interest for the general public except mathematicians; the Coinami \citep{ileri2016coinami} proposes a solution to solve DNA sequencing problems for PoUW, but it needs a centralized authority and is not genuinely decentralized; the CoinAI \citep{baldominos2019coinai} propose to develop a Proof-of-Learning system which uses the final performance as the certificate, but setting a reasonable ``performance bar'' to desired difficulty is a hard (if even possible) task.

In the specific area of Proof-of-Learning (PoL), \citet{jia2021proof} considered a setting of a specific \emph{threat model}, and proposed a PoL mechanism to show that the verification of SGD training requires two types of parties as \emph{provers} and \emph{verifiers}. %They aim to design a mechanism in which an honest certificate generated by the prover can be verified by the verifier at a low computational cost, while a dishonest certificate (spoof) \emph{within the threat model} will be detected by the verifier at a low cost too.
In their protocol, the provers report the state every $k$ epochs and the verifier checks the \emph{largest updates}, arguing that within their threat model, the largest updates tend to be the most suspicious when the dishonest prover attempts to forge a fake certificate. However, when going beyond that specific threat model, \citet{zhang2022adversarial} showed that attackers can maliciously design spoofs that bypass the largest-update verification and exploit the tolerance. Furthermore, \citet{fang2023proof} claimed that the Proof-of-Learning ``is more broken than you think'' by demonstrating that realizing the desired security requirements reduces to solving hard open problems in learning theory, so that a provably {Byzantine-}secure PoL is not possible to design until significant progress in further understanding in deep learning.

In an economic view, the difficulty in designing a cheap but secure verification protocol of PoL is conceptually related to \emph{Goodhart's Law}: ``When a measure becomes a target, it ceases to be a good measure'' \citep{goodhart1984problems}. Until further understanding of deep learning, no more efficient method has been found to verify the integrity of training than training it again. The work of \citet{jia2021proof}, to reduce the computational overhead of the verification, chose to identify ``most suspicious'' parts to verify, but when the criteria for suspicion are deterministically designed, there would constantly be risks that cheaters adversarially design attacks to bypass the criteria. Therefore, designing an efficient method to deterministically (or with high probability) catch all cheats in PoL is indeed faced with major difficulties.

In contrast, our research relaxes the security requirement to ``incentive-security'' in a game-theoretical setting: we do not need to prevent all attacks, but only need to prevent attacks from being ``worthy''. Intuitively, while all attacks are considered equal in Byzantine security, they may have different degrees of effects in the economic view. In our design, our mechanism detects attacks in a stochastic way and ``more severe'' attacks that potentially benefit the attackers more, would be caught with higher chances and lead to heavier expected penalties. In this way, our incentive-secure PoL design can manage to disincentivize rational agents from cheating.

\added{Additionally, a recent work by \citet{cao2025sedulity} has extended an earlier version of our work for detailed implementation on real-world blockchains. Particularly, \citet{cao2025sedulity} adopted our incentive-security paradigm and our framework of probabilistic verification with the CTF protocol without our multi-verifier design (as in Appendix~\ref{subsection:multi:verifier:mechanism}), and further addressed the issues of energy wasting from competition via systematic \emph{scheduling} and additionally considered the \emph{model theft} issue. While we consider the scheduling of the tasks as an independent interest and would like to \textbf{focus on the interaction between provers and verifiers of a specific task}, the model theft issue can actually be addressed by incorporating the provers' identities into the random seeds in the prover's protocol. The details are discussed in Section~\ref{sss:theft}.}

Another difference between the settings of \citet{jia2021proof} and our work is that: while the work of \citet{jia2021proof} mainly aims to prevent the spoof of a \emph{specific} PoL to protect the copyright of the model, we additionally prevent all spoofs that try to cheat the verifier and claim that the training is correctly done, getting the training reward. Hence, while our work adopts a relaxed notion of incentive security, it generally applies to a wider range of attacks. The details are discussed in Section~\ref{subsec:prg}. 

\vspace{-0.7em}

\subsection{Settings of Trusted or Untrusted Problem Providers in PoUW Protocols} \label{subsec:untrusted:problems}

\begin{small}

\begin{table*}[htb]
\vspace{-1em}
\centering
\begin{tabular}{|c|c|c|c|}
\hline
           & Cryptographic                                                             & Game-theoretic (existing)                                                             & Ours                                                                                       \\ \hline
Approach   & Zero-knowledge Proofs                                                     & Verification Games                                                                    & Verification Games                                                                         \\ \hline
Example    & zkML                                                                      & opML, PoSP                                                                                  & Incentive-Secure PoL                                                                       \\ \hline
Security   & Cryptographic                                                             & \begin{tabular}[c]{@{}c@{}}Mixed-Strategy Nash Eq. (with few cheaters)\end{tabular} & \begin{tabular}[c]{@{}c@{}}Pure-Strategy Nash Eq. (with no cheater)\end{tabular}         \\ \hline
Overhead   & High ($\ge$ 1000x)                                                        & Moderate ($\ge$ 1x)                                                                   & Low ($\lesssim$ 0.1x)                                                                          \\ \hline

Challenges & \begin{tabular}[c]{@{}c@{}}High overhead, low scalability\end{tabular} & Verifier's Dilemma                                                                    & \begin{tabular}[c]{@{}c@{}}Communication cost  (for huge models)\end{tabular} \\ \hline
\end{tabular}

\caption{Comparison of Trustworthy AI Protocols on Blockchain}\label{table:comparison}

\vspace{-3em}
\end{table*}
\end{small}

In the traditional PoW mechanism, e.g. in Bitcoin, the hash puzzle is automatically generated from the previous block and is unpredictable before the previous block is confirmed. However, in the paradigm of PoUW, the problem should come from real-world providers, so can be indeed predictable or even controllable. In particular, malicious parties can conduct the following attacks:

\vspace{-0.3em}

\begin{itemize}
    \item Known-model attack: submit a problem to which they already have a solution, and then submit the solution to claim the block.
    \item Model-stealing attack: submit a model trained by others (or based on it) and claim that they trained it on their own.
\end{itemize}

\vspace{-0.3em}

As far as we are concerned, most research in the literature of PoUW has not considered the credibility of the problems, i.e. implicitly assumed that the problems are \emph{credible} and focus on the prevention of spurious certificates. %Besides, Coinami \citep{ileri2016coinami} extensively discussed their system structure that depends on authority nodes and stated that their system is ``not completely decentralized'' and argued that it is necessary for usefulness; 
For example, while the work of \cite{jia2021proof} did not consider known-model attack, their solution to model-stealing attack is a chain-of-trust protocol that also relies on a sort of authorization.

Nevertheless, to build a robust blockchain system, we are motivated to design a mechanism in which both problem providers and provers can be \emph{untrusted} but are incentivized to behave honestly, which we call \emph{frontend-secure}. In consideration of frontend-security, \citet{ball2017proofs} proposed a PoUW mechanism based on Orthogonal Vectors that adds an extra randomization layer to the PoUW challenge: instead of only requiring the prover to solve the problem, it requires the prover to solve the problem ``in the way the system (randomly) specifies'', so that even if the prover has a solution beforehand, the transcript may not meet the requirement of the challenge and the prover still has to compute the challenge again to pass the verification. %The protocol works as follows:
\iffalse
\vspace{-0.5em}
\begin{itemize}
    \item The system receives the problem $A$ from an untrusted problem provider.
    \item The system generates a random seed $\phi$ and transform $A$ to a PoUW challenge $C=\mathcal{C}(A,\phi)$.
    \item The prover solves the challenge and gets a certificate $c=S(C)$.
    \item The verifier verifies the certificate, expecting to get $V(C,c)=true$.
    \item The system recovers the solution $w=\mathcal{W}(C,\phi)$ and sends it to the problem provider.
\end{itemize}

\vspace{-0.5em}
\fi

%On a high level, the frontend-security of the proposal is based on the one-way reduction from $C$ to $A$: it is easy to generate a solution to $A$ from a solution to $C$, but not in the inverse direction. 
While our design is generally different from this work, we indeed adopt the thought to introduce randomization in the design of PoUW challenges, which is naturally implementable due to the stochastic nature of the training of deep learning models.

\vspace{-0.7em}

\subsection{Trustworthy AI and MLaaS on the Blockchain Platform}

While the artificial intelligence (AI) has been becoming one of the most attractive topic in research and industry, the expansion of model sizes and computing source consumption in machine learning tasks has raised significant concerns about security \cite{bertino2021ai,hu2021artificial} and sustainability \cite{khakurel2018rise}. The advent of Machine Learning as a Service (MLaaS) \cite{ribeiro2015mlaas} has democratized access to powerful AI tools, enabling companies and individuals to integrate advanced machine learning models into their operations without extensive infrastructure. 

%However, this convenience comes with challenges in ensuring the transparency \cite{antunes2018fairness,von2021transparency} and security \cite{qayyum2020securing} of these services. Trustworthy AI principles are crucial in this context, as they advocate for the development and deployment of AI systems that are secure and accountable \cite{kaur2022trustworthy}. 

The blockchain, as a decentralized and transparent infrastructure, has an inherent affinity for applications in trustworthy AI \cite{nassar2020blockchain}. Furthermore, the innate element of cryptocurrency tokens can also serve as economic incentives for participation \cite{lee2019decentralized}. 

Three recent methodologies that implement trustworthy AI in the blockchain platform are zero-knowledge machine learning (zkML) \cite{zkml}, optimistic machine learning (opML) \cite{conway2024opml} and Proof-of-Sampling (PoSP) \cite{zhang2024proof}. The method of zkML utilizes the tool of zero-knowledge proof to secure the integrity of inference, but the nature of zero-knowledge proof makes the protocol extremely inefficient. The methods of opML and PoSP adopt economic incentives in the protocol and reduce the computational overheads to one or a few additional passes of computation, but opML effectively addresses the Verifier's Dilemma to prevent verifiers from being lazy when the fraction of dishonest provers is \emph{arbitrarily} low\footnote{It utilizes constant penalty that works when the fraction $\epsilon$ of dishonest provers is at least a small constant, but does not work uniformly when $\epsilon \to 0$.}, and the small challenging probability of PoSP leads to high staking requirements of verifiers and low detection probabilities of cheats, which undermine the user-friendliness and robustness of the protocol. In comparison, our mechanism has a computational overhead as low as a small fraction of one training pass, and it utilizes the capture-the-flag protocol to bypass the Verifier's Dilemma (See Section~\ref{subsec:verifier:strategy} and Theorem~\ref{thm:verifier:dilemma}) and prevent lazy verifiers robustly when there are arbitrarily few or no cheating provers. 
\added{Besides, \cite{zhao2024takes} proposed a peer-prediction methodology to evaluate verifiers' reports against each other when no centralized ground-truth is available, which inspired our design of the multi-verifier voting scheme, and also showed the \emph{compactness} criteria indicating that the high-penalty protocol in \cite{conway2024opml,zhang2024proof} may undermine robustness compared to our design.}
We show the comparison of the related protocols in Table~\ref{table:comparison}.

Hence, the family of Proof-of-Learning mechanisms, especially in the paradigm of Proof-of-Computation, not only serves as a fundamental mechanism to maintain the reliability of blockchain systems but also has the potential for the development of low-overhead decentralized computing power markets.

\vspace{-1em}
\section{Preliminaries}
\label{sec:prelim}
%In this section, we first introduce the general pipeline of the Proof-of-Learning protocol, and 

{
In the Proof-of-Learning mechanism, we consider a situation where a prover tries to convince all parties via a ``certificate'' that she has honestly completed the training task and is thus eligible to claim the block reward; the verifier, in turn, is expected to verify the validity of the certificate to ensure the security of the system. In general, our protocol works as follows:

\begin{enumerate}
\item A PoL problem $A$ is assigned.
\item One or more provers work on the problem $A$, either honestly or dishonestly, until one prover claims to have solved the problem and posts the PoL certificate $c$, winning the competition; other provers lose the competition and have their computing efforts lost as a sunk cost.
\item The verifier verifies the certificate $c$, possibly via interactions with the prover, and reports the verification result.
\item The system processes rewards and penalties accordingly.
\end{enumerate}

In the rest of this section, we briefly discuss the basic components of the protocol. %Then in Sections~\ref{section:overview}-\ref{sec:basicIS} we discuss about our design for the incentive-security for the prover, and in Section~\ref{sec:verIS} about an augmentation for the incentive-security for the verifier.
}

\vspace{-0.7em}

\subsection{Modeling of ML Training Tasks}

Suppose there is a data distribution $\D$ in the form of $\X\times\Y$, in which $\X$ is the input space and $\Y$ is the output space. A machine learning model (abbreviated as ``model'') is a function $f:\W\times \X \to \Y$ in which $\W$ is the parameter space. In the ML practice, the parameters are commonly called \emph{weights}.

The ML training task can be modeled as \emph{empirical risk minimization}, in which a training dataset is sampled from the distribution as $D_{tr}\sim \D^n$, and we denote $D_{tr}=(d_1,\cdots,d_n)$ in which $d_i = (x_i,y_i)$. For any data point $(x,y)$ and weight $w\in\W$, the model prediction is $f(w,x)$, and the loss is defined as a \emph{loss function} $\L(f(w,x),y)$. Then, the empirical risk to minimize is defined as:
\begin{equation}
    \hat{L}(w) = \sum_{i \in [n]} \L(f(w,x_i),y_i).
\end{equation}

\vspace{-0.7em}
The stochastic gradient descent (SGD) training process consists of a number $E$ of \emph{epochs}, and every epoch corresponds to one full pass of the training set. In each epoch $e\in [E]$, the training set is randomly divided into $l$ batches of size $m$, with $n=l\cdot m$. In every step $s=(e-1)m+j$, the corresponding batch, denoted as a subset $b_e(j)$ of $[n]$, is processed, and the weight is updated as:
\begin{equation}
    w_s = T_{\eta, b_e(j)}(w_{s-1}) = w_{s-1} - \eta \cdot \nabla{\hat{L}_{b_e(j)}(w_{s-1})}.
\end{equation}

Here, $\eta$ is a hyper-parameter of learning rate and $\hat{L}_{b_e(j)}$ is the empirical risk on the batch $b_e(j)$, defined as:
\begin{equation}
    \hat{L}_{b_e(j)}(w) = \sum_{i \in b_e(j)} \L(f(w,x_i),y_i).
\end{equation}

Therefore, given the batch division as $b_e\in \B$, the training process of epoch $e$ can be formulated as a mapping $\T_\eta: \B \times \W \to \W$, with
\begin{small}
\begin{equation}
    \T_\eta(b_e, w) = T_{\eta, b_e(m)}(T_{\eta, b_e(m-1)}(~~ \cdots~~ T_{\eta, b_e(1)}(w) ~~\cdots~~)).
\end{equation}
\end{small}

In the rest of this paper, we regard $\eta$ as a fixed hyper-parameter and denote $\T_\eta$ as $\T$ for simplicity.

\vspace{-0.5em}

\subsection{Credible (Pseudo-)Randomness Generator}\label{subsec:prg}

As described above, due to the random choice of batches $\{b_e(j)\}$, the training process $\T$ of \emph{\textbf{stochastic}} gradient descent, is innately a \emph{\textbf{stochastic process}}. To verify the correctness of the training process, the paper of \citet{jia2021proof} leverages the concentration properties of the process and introduces \emph{tolerance} for slight discrepancies in verification. However, the tolerance can in turn be exploited for adversarial attacks (See in \citep{zhang2022adversarial}).

In Bitcoin, the randomness in the hash puzzle is essentially based on a pseudo-randomness generator (cryptographic hash) seeded with the last block, so that every party can have a consensus on the same pseudo-random PoW challenge. 

%A typical pseudo-randomness generator (PRG) works as follows. Given a random seed $\phi$, the PRG generates a sequence of $r_\phi(1),r_\phi(2), \cdots$, and without loss of generality, we assume they are uniformly distributed in $[0,1)$. Since the PRG is typically based on a finite state machine, the sequence will eventually repeat after a period. Nevertheless, a ``good'' PRG would have a period long enough and pass certain randomness tests, and a PRG that meets the cryptographical criteria is called ``cryptographically secure'' \cite{kelsey1998cryptanalytic}.

In this paper, we would perform the SGD training with $\{b_e\}$ generated from a cryptographically secure PRG with seeds generated from the previous block, so that the prover and verifier will run with the same pseudo-random sequences and get exactly the same result for the same epoch. On the other hand, as the sequence is not predictable until the seed $\phi$ is generated, even if a strategic party submits a task with a known model and training process, as the protocol requires the prover to train with the given random seed, the prepared model or training process would not pass the verification and she still has to train it again to claim the reward.

\subsubsection{Model Theft Prevention} 
\label{sss:theft}

\added{It has been discussed by \citet{cao2025sedulity} in their extended implementation of an earlier version of our work that, if the random seed $\phi$ is given without consideration of the prover's identity, an attacker may take the trained model without acknowledging the prover's copyright. This issue can actually be addressed via encoding the prover's identity data into $\phi$. For example, we can set $\phi=Hash(prev\_block, prover\_id)$. Hence, even before the PoL is confirmed on the blockchain, the prover may show his $prover\_id$ and demonstrate that the random seed $\phi$ is bound to his unique identity, thus identifying the model theft attack in both on-chain and off-chain scenarios.}

\vspace{-0.5em}

\subsection{Modeling of Prover's Incentive}
\label{subsec:inc:model}
For a fixed prover and a fixed task, %as the training process is linear, 
we can assume the computational cost to honestly train an epoch is a deterministic constant $m$, and thus honestly training the task has a cost (aka.~``difficulty'') of $M=m\cdot E>0$, which can be dynamically adjusted by adjustment of $E$. For each epoch, the prover may train it honestly or dishonestly (detailed discussion in Section~\ref{sec:basicIS}). When dishonestly training an epoch, the prover may pay a significantly lower computational cost, and we assume it to be $0$. We assume that dishonest training of one epoch does not affect the computational cost of further epochs. Therefore, if we honestly train {a $\rho$ portion of all epochs, the computational cost is (lower bounded by) $\rho M$.} 

There can be competition among provers (or not, due to the allocation rule of the tasks) {and only the first prover who submits a certificate wins, so if a prover does more honest computation and consumes more time before submission, her probability of winning the competition does not increase.} We define $P:[0,1]\to (0,1]$ as a {non-increasing} function that characterizes the competition: if the prover computes $\rho$ portion of the task (i.e.~$\rho E$ epochs) honestly, then she has a $P(\rho)$ probability of winning, in which $P(0)=1$. If there is no competition, we just let $P(x)\equiv 1$.

When the prover wins the competition and submits her certificate, if $\rho<1$, i.e.,~the prover does not act honestly, then there is a chance that she is caught. For any fixed $\rho$, {as the prover may have multiple strategies to choose the $(1-\rho)$ portion for cheating}, we denote $Q(\rho)$ as the maximal probability {among all such cheating strategies} of passing the verification, {in which we assume $Q(\cdot)$ is monotonic non-decreasing and $Q(1)=1$}. If passing the verification, the prover gets a reward of $R$ {at a computational cost of $\rho M$}, and the net utility is $R-\rho M$; if getting caught cheating, she will be penalized for $\gamma R$, and the net utility is $-(\gamma R + \rho M)$. For a good PoL mechanism, we expect a low $\gamma$, ideally zero, to lower the staking requirement{\footnote{To ensure that the prover has enough tokens to pay the penalty, we have to require the prover to stake $\gamma R$ before participation. We can see that setting $\gamma\to+\infty$ makes the problem trivial as the prover gets an infinite penalty whenever she cheats; however, it needs the prover to stake an infinite amount of tokens, which is not possible.}} {and improve the convenience} of participation.

If the prover loses the competition, the sunk cost in training the model is still paid, but she may find out that the task has been completed by another prover before she completes the computation, {so the cost can be less than $\rho M$.} Hence, we denote her expected utility {conditioned on losing} as $-\mu(\rho) \in [-\rho M, 0]$. Assuming $P(\cdot)$ is a differentiable function, we can compute that (details in Appendix~\ref{app:sunk:cost}):
\begin{equation}
\mu(\rho) = \frac{\int_0^{\rho} P(x) dx - \rho P(\rho)}{1-P(\rho)} M.
\label{eqn:sunk:cost}
\end{equation}

In summary, the expected utility for the prover to honestly train a $\rho$ portion of the task is
\begin{small}
\begin{equation}
\begin{aligned}
\label{eqn:utility}
    u(\rho) &= P(\rho)(Q(\rho)\cdot (R-\rho M) - (1-Q(\rho))\cdot (\gamma R+ \rho M)) \\
    &\qquad -  (1-P(\rho)) \mu(\rho)  \\
    &= P(\rho)(Q(\rho) - \gamma (1-Q(\rho)))R -   \int_0^{\rho} P(x) dx \cdot M .
\end{aligned}
\end{equation}
\end{small}

To make the mechanism desirable for the prover and incentivize the prover to honestly train all the $E$ epochs, we expect to satisfy the following (strict) interim individual-rationality (strictly interim IR) and basic incentive-security (BIS) properties:

\begin{definition}[Strict interim individual-rationality]\label{def:ir}
    We call a PoL mechanism strictly interim individually-rational (strictly interim IR) if and only if honestly training the task earns a positive expected utility, i.e.,~    \begin{equation}
        u(1)>0,
    \end{equation}
    assuming the verifier is honest.
\end{definition}

\begin{definition}[Strict interim basic incentive-security]\label{def:bis}
    We call a PoL mechanism strictly interim basic incentive-secure (strictly interim BIS) if and only if honestly training the task earns strictly more expected utility than dishonest training, i.e.,~
    \begin{equation}
        \forall \rho \in [0,1), u(\rho)<u(1),
    \end{equation}
    assuming the verifier is honest.
\end{definition}

In the rest of this paper, without confusion, we omit the words ``strict'' and ``interim'', and call a mechanism $\gamma$-IR-BIS if it satisfies both of the properties above for parameter $\gamma$.

\subsection{Threat Model}
\label{subsec:threat:model}

\subsubsection{Provers}

$~$

In light of dishonest provers, \citet{jia2021proof} introduce a threat model that consists of 4 types of attacks, as follows:

\begin{itemize}
\item Retraining-based spoofing: the attacker aims to get the same PoL of the same model.
\item Stochastic spoofing: the attacker aims to get a different PoL of the same model.
\item Structurally correct spoofing: the attacker aims to get an invalid PoL of the same model that passes verification.
\item Distillation-based spoofing: the attacker aims to get a PoL of a (slightly) different model.
\end{itemize}

While our mechanism has some structural similarity to \citep{jia2021proof}, our work has a different motivation. The work of \citet{jia2021proof} mainly aims to protect the copyright of an already trained \emph{model}, but in our work the PoL serves as a Proof-of-Useful-Work, and our mechanism additionally aims to verify that the prover (as a miner) honestly did the computation. 
%in which the attacker may have the interest to steal the copyright or not.
\added{In our work, we alternatively address the model theft attack via incorporating the prover's identity into the random seed, as shown in Section~\ref{sss:theft}.}

\ifdefined\EnableAIS{In this sense, we introduce two classes of dishonest attackers:

\begin{itemize}
\item Rational attacker: the attacker aims to generate a PoL that passes the verification with less computational cost, whether the model is the desired one or not.
\item Malicious attacker: the attacker aims to generate a PoL of a significantly incorrect model that passes the verification to mislead the model's users, possibly at a higher cost.
\end{itemize}
}
\else{
Nevertheless, as PoW miners typically compete for the blocks to earn \emph{block rewards}, so we are motivated to mainly consider \emph{rational} miners who would cheat to gain more economic utility.
}

In the paper of \citet{jia2021proof}, the authors assume the attacker has the full information of the desired model, the full dataset, but does not have information of the random source of the model. In our paper, as the random seed is specified by the protocol, we consider an \emph{even stronger} adversary that also has the random source. Formally, we assume that:

\begin{itemize}
\item The attacker has full information of the desired model $f(W,\cdot)$ trained with seed $\phi$, but does not know the training process (for model theft attacks); she has also pre-trained a valid model $f(W',\cdot)$ with a different seed $\phi'$ (for known-model attacks).
\item The attacker has full information on the dataset.
\item The attacker also has the random source of the desired model, i.e. the random seed $\phi$ and the randomization guideline $\G$.
\end{itemize}

{With our rational attacker assumption, the attack space contains a slightly modified version of 4 types of attacks.} Actually, it is stronger because the structurally correct spoofing no longer requires to get the same model.

\begin{itemize}
\item Retraining-based spoofing: the attacker aims to get the same PoL of the desired model $f(W,\cdot)$.
\item Stochastic spoofing: the attacker aims to get a different but valid PoL of the desired model $f(W,\cdot)$.
\item Structurally correct spoofing: the attacker aims to get an invalid PoL of any (correct or incorrect) model $f(W^\#,\cdot)$ that passes verification.
\item Distillation-based spoofing: the attacker aims to get a valid PoL of a (slightly) different model $f(W'',\cdot)$.
\end{itemize}

In Section~\ref{sec:basicIS} we will show the incentive-security property of our basic and full mechanisms against such attacks.%Further in Section~\ref{sec:advancedIS}, we additionally discuss about an augmentation of our mechanism to satisfy adversarial incentive-security against malicious attackers who can particularly conduct structurally correct spoofing and gain additional utility based on $\|W^\#-W\|$.

\subsubsection{\added{Verifiers}}

$~$

%\color{blue}

While previous studies on PoL (e.g., \cite{jia2021proof}) usually assume honest verifiers, in real-world applications, the verifiers can also be strategic players. Furthermore, this issue becomes more prominent particularly for computation-intensive verification schemes, due to the infamous ``Verifier's Dilemma'' (see  \cite{luu2015demystifying,zhao2024takes} and Section~\ref{subsec:verifier:strategy}) in which rational verifiers are incentivized to lazily accept the proof. To model the verifiers' actions, we categorize the verifiers' behavior into three classes:

\begin{itemize}
    \item Honest: to honestly verify the proof and report as in the protocol.
    \item Lazy: to report some (forged) information without verifying the proof.
    \item Malicious: to verify the proof but report differently than the protocol specifies.
\end{itemize}

We regard both lazy and malicious behaviors as ``dishonest'', but they have inherent differences. For example, a malicious verifier still needs to incur the verification cost, but a lazy verifier does not need to. We will discuss the Verifier's Dilemma in detail in Section~\ref{subsec:verifier:strategy}, and our verifiers' incentive guarantees in Section~\ref{subsec:vis}.

%\color{black}
\section{Basic Mechanism for Trusted Verifiers}
\label{section:overview}
In this section, we provide a general overview of our basic protocol for provers and verifiers, under the assumption of trusted verifiers which is widely adopted in previous literature.

\subsection{Generation of PoL Certificate}
\label{subsec:prover:prot}

The protocol is shown in Algorithm~\ref{alg:p1}. For each block, we assume that there is an assigned problem $A=(D_{tr},\E,\phi)$, in which $D_{tr}$ is the training dataset, $\E$ is the environmental variables which include learning rate $\eta$, loss function $\L$, batch size $m$, number of epochs $E$, randomization guideline $\G$, \added{model $f$, recording interval $k$, and the verification parameter $\alpha$}, which dictates how the randomness is generated from the seed, and other required specifications if needed (e.g. the initialization), and $\phi=Hash(prev\_block, prover\_id)$ is the random seed generated from past blocks and the prover's identity to prevent model theft attack (as discussed in ~\ref{sss:theft}).

The prover is expected to solve the problem $A$ by training $E$ epochs following the given rule directed by $\E$, with the random seed $\phi$. The initialization $w_0$ is specified by $\E$, and the prover is required to record the status after every $k$ epochs, in which $k$ is an integer parameter (either specified in the blockchain rule or specified in $\G$): smaller $k$ leads to larger certificate size and prover storage consumption but lower computational overhead (see Section~\ref{sec:basicIS}).

We assume that $E$ is divisible by $k$, then the training process can consist of $T=\frac{E}{k}$ \emph{stages}, in which each stage consists of $\tau = k\cdot l$ steps. For each stage $t \in [T]$, the prover is required to save the current weight $W_t = w_{t\cdot \tau}$.
%and the (full-dataset) empirical risk $\hat{L}_t = \hat{L}(W_t)$
To save on-chain space, we only need the prover to a hash value of each $W_t$, and the required certificate is structured as $c=(c_1,\cdots,c_T)$ in which $c_{t} = hash(W_t)$; In the verification stage, she also needs to post a subset of $\{W_t\}$ when queried by the verifier (see section~\ref{subsec:verifier:prot}).

Denote $|\W|$ as the model size, then the communication complexity is $O(\frac{E}{k})$ and the storage requirement for the prover is $O(\frac{E|\W|}{k})$ on this part. 
\begin{small}
\begin{algorithm}[htb]
\caption{Prover's certificate generation protocol in the basic mechanism}
\label{alg:p1}
\begin{algorithmic}[1]
%\Procedure{Euclid}{$a,b$}\Comment{The g.c.d. of a and b}
%\State Construct new vertices $I,O$.
%\State $V=S\cup\{I,O\}$
%\State $E=\emptyset$
\State Input $A=(D_{tr},\E,\phi)$.
\State Initialize $w=w_0$ according to $\E$.
\State $T:=\frac{E}{k}$.
\State $e:=0$
\For {$t:= 1\cdots T$}
    \For {$x := 1 \cdots k$}
        \State $e:=e+1$
        \State Draw $b_e$ according to $(\G,\phi)$
        \State $w_{(e)}:=\T_\eta(b_e,w_{(e-1)})$
    \EndFor
    \State $W_t:=w_{(e)}$
    %\State $\hat{L}(W_t):= \sum_{i} \L(f(W_t,x_i),y_i)$
    \State $c_t:= hash(W_t)$
\EndFor
\State Post $c:=(c_1,\cdots,c_T)$.
\end{algorithmic}

\end{algorithm}
\end{small}

\subsection{Verification}
\label{subsec:verifier:prot}

The verification protocol is shown in Algorithm~\ref{alg:v1}. The verifier is expected to \emph{randomly}\footnote{In this paper, whenever we use the term ``randomly'', we refer to ``randomly with a uniform distribution'' unless specified otherwise.} verify $\alpha$ stages ${\rm{t}}_{ve}=\{t_1,\cdots, t_\alpha\}$ among $T$, in which $\alpha$ is a security parameter. For unpredictability to the prover, these stages should be drawn via uniform random sampling without replacement {from her own secret} (independent from $\phi$). Then the verifier posts ${\rm{t}}_{ve}$, {requiring} the prover to show corresponding weights. 

Then, for each $t_i$, the prover is expected to post the weights before and after the stage, i.e. $W_{t_{i}-1}$ and $W_{t_i}$. The verifier then checks whether the previously posted hashes are correct, and re-train the stage from $W_{t_{i}-1}$ to see if the result is $W_{t_i}$. If and only if all tests are passed, then the basic verification is successful; otherwise, the verifier reports the detected cheating stages and indicates that the verification has failed. 

\ifdefined\enableAIS
To ensure adversarial incentive-security (as defined in Section~4.3), the verifier runs \emph{anomaly detection}, and can require to check additional stages if necessary. The detailed discussion is deferred to Section~\ref{sec:advancedIS}.
\else{}\fi

In this part, the communication complexity is $O(\alpha|\W|)$ and the relative computational overhead is $O(\frac{\alpha k}{E})$. In total, the communication complexity is $O(E+\alpha |\W|)$.

\begin{small}
\begin{algorithm}[htb]
\caption{Verifier's verification protocol in the basic mechanism}
\label{alg:v1}
\begin{algorithmic}[1]
%\Procedure{Euclid}{$a,b$}\Comment{The g.c.d. of a and b}
%\State Construct new vertices $I,O$.
%\State $V=S\cup\{I,O\}$
%\State $E=\emptyset$
\State Input $A=(D_{tr},\E,\phi), c=(c_1,\cdots,c_T)$.
\State Draw $\t_{ve}=\{t_1,\cdots,t_\alpha\}$ from $\{1,\cdots,T\}$ via her own secret.
\State Post $\t_{ve}$ to the prover, expecting to get $\{(W_{t_i-1},W_{t_i})\}$ for each $t_i \in \t_{ve}$.
\For{$i\in 1\cdots \alpha$}
    \If {$c_{t_i-1} \neq hash(W_{t_i-1}) \vee c_{t_i} \neq hash(W_{t_i})$}
        \State Return (``Fail'', InvalidWeights($t_i$))
    \EndIf
    \State $w = W_{t_i-1}$
    \For {$e := k\cdot (t_i-1)+1,~ \cdots,~ k\cdot t_i$}
        \State Draw $b_e$ according to $(\G,\phi)$
        \State $w:=\T_\eta(b_e,w)$
    \EndFor
    \If {$w\neq W_{t_i}$}
        \State Return (``Fail'', ErrorInStage($t_i$))
    \EndIf
\EndFor
\State Return ``Success''
\end{algorithmic}

\end{algorithm}

\end{small}

%\subsection{Full Mechanism for Untrusted Verifiers}

\section{Full Mechanism for Untrusted Verifiers}
\label{sec:verIS}

In this section, we discuss the verifier's incentive and augment our design to incentivize the verifier to verify honestly. On a high level, we introduce \emph{safe deviations} as ``flags'' that do not affect the validity of the PoL but gain the verifier additional rewards that compensate for the verification cost, and design economic incentives to incentivize the verifier to find as many flags as possible within the $\alpha$ stages they inquire for their optimal utility, so that they would indeed verify $\alpha$ stages as supposed to.

\subsection{The Verifier's Dilemma}
\label{subsec:verifier:strategy}
In the previous works on Proof-of-Learning, it is typical that the systems only prevent the provers from cheating while assuming that verifiers are honest. However, in a fully decentralized and permissionless blockchain system, this is not necessarily true. While one may straightforwardly consider game-theoretic ways to incentivize verifiers to verify honestly, the Verifier's Dilemma \cite{smuseva2022verifier,fiz2020sluggish} would occur:

\begin{framed}
\begin{center}
\begin{large}
\textbf{\emph{Verifier's Dilemma}}
\end{large}
\end{center}
\begin{itemize}
\item If a PoUW mechanism is (incentive-)secure against strategic provers, then no (rational) prover would cheat.
\item If no prover would cheat and the verification has a non-zero computational cost, then the verifier's optimal strategy is to report ``Success'' without verification.
\item If all verifiers are rational and would not actually verify, then the security properties no longer hold.
%\item Therefore, it is difficult, if not impossible, to design a mechanism, in which the verifier simply reports ``Success'' or ``Fail'', with a \emph{Nash equilibrium} that both the prover and verifier act honestly.
\end{itemize}

\end{framed}

The Verifier's Dilemma indicates the difficulty in the design of a truthful mechanism with a \emph{Nash equilibrium}%\footnote{A Nash equilibrium \cite{nash1951non} refers to a situation in multi-party games in which no single party can benefit from individual deviation.} 
that both the prover and verifier act honestly. 

Formally, we can model the verification game as follows:

\begin{definition}[Verification Game]
In a verification game, there is one prover $P$ and $n_v\ge 1$ verifier(s) $V_1,\cdots,V_{n_v}$. The prover has an action space $A_p$, and a subset $A_p^H\subseteq A_p$ is denoted as \emph{honest}. We denote $A_p^D = A_p \backslash A_p^H$ as the set of the prover's dishonest actions. For each action $a_p\in A_p$, the prover is incurred an initial cost $c_p(a_p)$.

We assume $n_v$ verifiers are independent and homogeneous. Any verifier also has an action space $A_v$ with subsets $A_v^H$ and $A_v^D$ defined similarly. For any action $a_v\in A_v$, the verifier pays a cost of $c_v(a_v, a_p)$ and observes a result ``Success'' or ``Fail'', possibly attached with additional information in $\I$. Here, we denote $P_v(a_v,a_p)$ as the probability that the result is ``Success''.

In this work, we assume that the honest verification process may fail to detect cheats, but always passes honest proofs, i.e.,
\[
    a_p\in A_p^H \wedge a_v\in A_v^H \implies P_v(a_v,a_p) = 1.
\]
Finally, the prover and verifiers are rewarded or punished based on the verifiers' reports and the prover's action, given that the prover may dispute and future users may check the verification result and do \emph{slashing} for dishonest verification. Hence, the payment rule can be denoted as: 
\[
\pi: (\{\textup{``Success'', ``Fail''}\}\times \I)^{n_v} \times A_p \to \mathbb{R}^{n_v+1}. 
\]
%For the slashing rule, since the honest verification always passes honest proofs, we assume that reporting ``Fail'' when $a_p\in A_p^H$ can be regarded as \textbf{deliberately malicious} and will incur heavy penalties $(\to \infty)$ for the verifier.

\end{definition}

From the modeling, we can show a formal negative result as:

\begin{theorem}[Verifier's Dilemma]
\label{thm:verifier:dilemma}
In a verification game in which the only information the verifier(s) report is ``Success'' or ``Fail'', i.e. $|\I|=1$, and honest verification has a strictly positive cost, i.e. 
\[
    a_v\in A_v^H \implies c_v(a_v, a_p)>0,
\]
it is impossible to design a verification mechanism with a pure-strategy Nash equilibrium that the prover and verifier(s) simultaneously act honestly.
\end{theorem}
The proof is deferred to Appendix~\ref{app:proof:dilemma}.

\textbf{Failure of the basic mechanism.} While we may straightforwardly want to reward the verifier for catching cheats, like the idea in \cite{conway2024opml}, unfortunately, from the Verifier's Dilemma, as long as the reward for the verifier is bounded, we can see that our basic mechanism in Section~\ref{section:overview} would not work. Formally, we have

\begin{theorem}\label{thm:veri:neg}
    In our basic mechanism in Section~\ref{section:overview}, if we assume that the verifier's maximum reward for finding a cheat is $v_+$ and the verifier's expected reward when the PoL passes the verification is $v_0$, then if $v_+\le v_0$ or $\epsilon \in (0, \frac{M}{T(v_+-v_0)})$, the verifier's strictly optimal strategy is to report ``Success'' without verification, i.e., to be lazy.
\end{theorem}

The proof of Theorem~\ref{thm:veri:neg} is deferred to Appendix~\ref{app:veri:neg}. Therefore, for any fixed $v_+, v_0$, we always have $\epsilon>0$ which makes the mechanism not incentive-secure for the verifier, because for $\epsilon$ small enough, the expected ``additional reward'' for catching a cheat would not cover the cost of verification. 

We empirically show the effects of the Verifier's Dilemma in Appendix~\ref{app:vreward}. 
In such experiments, we observe that even if there is a substantial fraction of dishonest provers, the verifiers may still be incentivized to verify less stages than required with a \emph{partially lazy} behavior, undermining the security of the protocol.

Therefore, we aim to modify the basic mechanism so that the verifier maximizes her expected utility by verifying and reporting honestly, uniformly for any sufficiently small $\epsilon$.
In this setting, we define verifier incentive security ($\epsilon$-VIS) as follows:
\begin{definition}[Verifier incentive security]
\label{def:vis}
We call a PoL mechanism $\epsilon$-VIS if and only if, as long as the prover is honest with a probability greater than $1-\epsilon$, the verifier gets a positive expected utility for a stage if and only if she verifies the stage honestly. \footnote{In our design, the verifiers' rewards are given stage-wise, as shown in Section~\ref{subsection:multi:verifier:mechanism}.}

%In simplicity, we call a mechanism VIS if it is $\epsilon$-VIS for some $\epsilon>0$.
\end{definition}

\subsection{The Capture-The-Flag Protocol}
\label{subsec:CTF}

As discussed in the parts above, we are aware that the Verifier's Dilemma only occurs in the scenario of $\epsilon\to 0$. Hence, a natural idea is to increase $\epsilon$, i.e. insert deliberate invalid objects, or so-called ``flags'' to incentivize verifiers to find, as in the works of \citep{reynouard2024bar,teutsch2024scalable,luu2015demystifying}. On the other hand, our Theorem~\ref{thm:verifier:dilemma} also shows the necessity for a desirable verification mechanism to let the verifier incorporate additional information into her report. Hence, the most straightforward idea is to deliberately generate invalid PoL's into the pool that serve as flags. However, this approach also faces the following challenges:

\begin{itemize}
    \item The cheaters in the pool can have complicated behavior, e.g., having different $\rho$'s in their cheating patterns. It is difficult to set proper $\rho$'s or analyze verifiers' behavior in the presence of both cheats and deliberately inserted flags.
    \item Particularly, if $\rho$ is not close to $0$, then the generation of invalid PoL's needs to contain a large portion of honest computation which has immense computational overhead, which not only undermines the efficiency but also complicates the protocol, e.g., in the allocation and compensation of such ``chores''.
    \item If $\rho$ is close to $0$, then the verifier would have a high probability of identifying the flags even if they only verify $1$ stage (rather than $\alpha$), which could incentivize a different dishonest strategy rather than the honest one.
\end{itemize}

In consideration of the issues above, we propose a variant to (let provers) insert the flags into each PoL certificate, i.e. designate a random subset of the stages as flags, and provers should make commitments about the flags inserted when submitting the PoL. However, due to the sequential nature of the SGD algorithm, inserting an invalid stage may affect the validity of the following stages and ultimately the resulting model; therefore, we insert \emph{safe deviations} that serve as flags, which is implemented by computing honestly with a differently designated seed. In particular, given the (root) random seed $\phi$, a stage $t$ can have 4 possible types:

\begin{enumerate}
\item Normal: it is trained with random seed $r_\phi(3t)$, as defined in Section~\ref{subsec:prg}.
\item Flag $F_1$: it is trained with random seed $r_\phi(3t+1)$.
\item Flag $F_2$: it is trained with random seed $r_\phi(3t+2)$.
\item Dishonest: otherwise.
\end{enumerate}

Notice that we do need two types of flags so that the verifier would be willing to check the type of the flag, instead of reporting ``Flag'' when the verification of ``Normal'' fails without any attempt to differentiate it from a dishonest stage. In this setting, we assume that less than half of the stages are flagged, so that the verifier would first verify with seed $r_\phi(3t)$ for stage $t$. If the verification of seed $r_\phi(3t)$ fails, the verifier, who believes that the probability of cheating is sufficiently small, would believe that it is a flag and randomly choose one of the following actions:

\begin{itemize}
    \item Verify with seed $r_\phi(3t+1)$. If successful report $F_1$, otherwise report $F_2$.\footnote{Since the stage is neither normal or $F_1$, it is either $F_2$ or dishonest. As the probability of cheating is sufficiently small, she would prefer to believe it is $F_2$ rather than take additional computational cost to distinguish them via verifying with seed $r_\phi(3t+2)$. Similar for the other case.}
    \item Verify with seed $r_\phi(3t+2)$. If successful report $F_2$, otherwise report $F_1$.
\end{itemize}

The verifier can alternatively randomly guess $F_1$ or $F_2$ without verification of either, but this would lead to a $\frac{1}{2}$ probability of reporting the wrong flag and getting penalized (for a higher amount than the flag reward). Hence, the verifier is incentivized to perform the verification as described above.

Therefore, if a cheater wants to disguise a dishonest stage as a flag, she must claim that it is $F_1$ or $F_2$ in the commitment, with a $\kappa=1/2$ probability of being caught
%\footnote{A stage fails the verification when the verifier reports a different type from the prover's commitment.}
if the stage is verified.

%\subsubsection{Toy Model: No Collusion}

%\added{To begin with the analysis, we first make an assumption that there is no collusion between the prover and the verifier. In this setting, we aim to achieve an honest \emph{Nash equilibrium} between the prover and the verifier, i.e., assuming one of them is honest, it is utility-optimal for the other party to also be honest. We consider the more challenging case of collusion in Section~\ref{sss:pp}.}

Intuitively, to incentivize the verifier to verify $\alpha$ stages among the total $T$, assume that we would like the prover to insert $\eta T$ (committed) flags in which $\eta\in [\frac{2\alpha}{T},\frac{1}{2})$, then when the verifier verifies honestly, the expected number of flags she finds would be $\alpha \eta$. Since the verifier only has access to the $\alpha$ stages in $\mathrm{t}_{ve}$, we would like to incentivize the verifier to find as many flags as possible so that the verifier would honestly verify all the $\alpha$ stages. Therefore, we award the verifier for each flag she detected. Particularly, recalling that the training cost of a stage is $\frac{M}{T}$ and noting that the discovery of a flag would take an additional $\frac{M}{T}$ cost of computation, we set positive parameters $ R_1 > \frac{M}{T}(\frac{2}{\eta}+1)$. \added{Then, assuming both the prover and the verifier are honest,} the system expects to give the verifier a reward of $W_v(u)$:
\begin{equation}
W_v(u) = R_1 u.
\end{equation}
\added{Nevertheless, the critical question occurs when the verifier detects a flag ($F_1$ or $F_2$) different from the prover's commitment. It may be because the prover is dishonest, but it may also be because the verifier is dishonest. If we only have one verifier for each task, we may not be able to decide unless we invoke additional committee voting which leads to complication.}

\added{To reduce the number of interaction rounds, we may embed the voting into the verification process itself, i.e., involving multiple verifiers in the verification process and rewarding verifiers based on the comparison of their reports with one another, following the idea of \cite{zhao2024takes}. Assuming the majority of verifiers are honest, we can hopefully resolve the disagreement using aggregated reports from the verifiers. We defer the detailed discussion of the multi-verifier voting protocol in Appendix~\ref{subsection:multi:verifier:mechanism}.}

\begin{small}
\begin{algorithm}[htb]
\caption{The pseudo-code of the full mechanism}
\label{alg:full}
\begin{algorithmic}[1]
%\Procedure{Euclid}{$a,b$}\Comment{The g.c.d. of a and b}
%\State Construct new vertices $I,O$.
%\State $V=S\cup\{I,O\}$
%\State $E=\emptyset$
\State Receive the task $A=(D_{tr},\E,\phi), \eta$, waiting for the prover to train.
\State The prover trains the model and posts the PoL certificate, posting $(c,\H)$ as in Algorithm~\ref{alg:p2}.
\State Collecting $n$ verifiers to join with their proposed $x_v$'s, reaching consensus on $\t_{ve}=\{t_1,\cdots,t_\alpha\}$ as stages to verify, as in Algorithm~\ref{alg:v2}, lines 1-3. 
\State The prover posts $\{(W_{t_i-1},W_{t_i})\}$ for each $t_i \in \t_{ve}$.
\State Each verifier $v_j$ independently verifies the PoL, posting commitment $hash(V^{(j)})$, as in Algorithm~\ref{alg:v2}, lines 5-28.
\State After all commitments are received, verifiers reveal $\{V^{(j)}\}$ and the prover reveals $\sigma$ as information of inserted flags.
\State Resolve the decision and rewards as in Section~\ref{subsection:multi:verifier:mechanism}.

\end{algorithmic}

\end{algorithm}
\end{small}

\begin{small}

\begin{algorithm}[htb]
\caption{Prover's certificate generation protocol in the full mechanism}
\label{alg:p2}
\begin{algorithmic}[1]

\State Input $A=(D_{tr},\E,\phi), \eta$.
\State Initialize $w=w_0$ according to $\E$.
\State $T:=\frac{E}{k}$
\State $e:=0$
\State Generate $\sigma = (\sigma_1,\cdots,\sigma_T)$ as a random permutation of $[T]$ from her own secret.
\State $\H:=hash(\sigma)$
\For {$t:= 1\cdots T$}
    \If {$\sigma_t \le \eta T$}
        \State \textbf{if} {$\sigma_t$ is odd~} \textbf{then}
           $s_t := r_\phi (3t+1)~$
        \textbf{else}
        $s_t := r_\phi (3t+2)$
    \Else
        \State $s_t := r_\phi (3t)$
    \EndIf
    \For {$x := 1 \cdots k$}
        \State $e:=e+1$
        \State Draw $b_e$ according to $(\G,s_t)$, denoted as $b_e := B_e(s_t)$.
        \State $w_{(e)}:=\T_\eta(b_e,w_{(e-1)})$
    \EndFor
    \State $W_t:=w_{(e)}$
    %\State $\hat{L}(W_t):= \sum_{i} \L(f(W_t,x_i),y_i)$
    \State $c_t:= hash(W_t)$
\EndFor
\State $c:=(c_1,\cdots,c_T)$ 
\State Post $(c,\H)$.

\end{algorithmic}

\end{algorithm}
\end{small}
\begin{small}
\begin{algorithm}[htb]
\caption{Verifier $v_j$'s verification protocol in the full mechanism}
\label{alg:v2}
\begin{algorithmic}[1]
%\Procedure{Euclid}{$a,b$}\Comment{The g.c.d. of a and b}
%\State Construct new vertices $I,O$.
%\State $V=S\cup\{I,O\}$
%\State $E=\emptyset$
\State Input $A=(D_{tr},\E,\phi), c=(c_1,\cdots,c_T)$.
%\State Draw $\t_{ve}=\{t_1,\cdots,t_\alpha\}$ from $\{1,\cdots,T\}$ via her own secret.
\State Broadcast a random number $x_{v_j}$ from her own randomness.
\State After collecting all $x_v$'s, draw $\t_{ve}=\{t_1,\cdots,t_\alpha\}$ from $\{1,\cdots,T\}$ with a random seed of $\{x_v\}_{v\in \{v_1,\cdots,v_n\}}$
\State Waits for the prover to post $\{(W_{t_i-1},W_{t_i})\}$ for each $t_i \in \t_{ve}$.
\For{$i\in 1\cdots \alpha$}
    \If {$c_{t_i-1} \neq hash(W_{t_i-1}) \vee c_{t_i} \neq hash(W_{t_i})$}
        \State $V_i:=$``Invalid''
        \State \textbf{continue}
    \EndIf
    \State $w = W_{t_i-1}$
    \State $w_1 = w$
    \For {$e := k\cdot (t_i-1)+1,~ \cdots,~ k\cdot t_i$}
        \State $b_e^{(0)}=B_e(r_\phi(3t))$
        \State $b_e^{(1)}=B_e(r_\phi(3t+1))$
        \State $b_e^{(2)}=B_e(r_\phi(3t+2))$
        \State $w_1:=\T_\eta(b_e^{(0)},w_1)$
    \EndFor
    \If {$w_1 = W_{t_i}$}
        \State $V_i := 0$
    \Else
        \State Draw $\xi \sim Uniform\{0,1\}$
        \If {$\xi = 1$}
            \For {$e := k\cdot (t_i-1)+1,~ \cdots,~ k\cdot t_i$}
                \State $w:=\T_\eta(b_e^{(1)},w)$
            \EndFor
            \State \textbf{if} {$w = W_{t_i}$~} \textbf{then~} $V_i := 1~$ \textbf{else~} $V_i := 2$
        \Else
            \For {$e := k\cdot (t_i-1)+1,~ \cdots,~ k\cdot t_i$}
                \State $w:=\T_\eta(b_e^{(2)},w)$
            \EndFor
            \State \textbf{if} {$w = W_{t_i}$~} \textbf{then~} $V_i := 2~$ \textbf{else~} $V_i := 1$
        \EndIf
    \EndIf
\EndFor
\State Post commitment $hash(V^{(j)}:=\{V_i\}_{i\in [\alpha]})$.
\State After all commitments are collected, reveal $V^{(j)}$.
%\If {$hash(\sigma)\neq \H$}
    %\State Return (``Fail'', InvalidFlagCommitment)
%\EndIf
%\For{$i\in 1\cdots \alpha$}
    %\If {$\sigma_{t_i} \le \eta T$}
    %    \State \textbf{if} {$\sigma_{t_i}$ is odd~}
    %    \textbf{then~} $s_i := 1~$ \textbf{else~} $s_i := 2$
    %\Else
    %    \State $s_i := 0$
    %\EndIf
    %\If {$V_i\neq s_i$}
    %    \State Return(``Fail'', ErrorInStage($t_i$) )
    %\EndIf
%\EndFor
%\State Return (``Success'', $\mathrm{t}_{ve}$, $V$).
\end{algorithmic}

\end{algorithm}
\end{small}

%\begin{small}

%\vspace{-2em}

%In Section~\ref{sec:basicIS}, we prove that for values of $\alpha, \beta, T$ that satisfy certain conditions, there is a Nash equilibrium that the prover trains honestly, and the verifier verifies exactly $\alpha$ stages.%, and in Section~\ref{sec:exp} we empirically show the result.}

%\textcolor{blue}{[Theoretical proof might be possible but I don't have an idea yet. The entire Section is not theoretically rigorous anyway.]}

%\subsubsection{Full Model: Possible Collusion}
%\label{sss:pp}

\section{Theoretical Incentive Security Analysis}
\label{sec:basicIS}

%\revise{[Can extend to full mechanism, adding verifier IS properties.]}

In this section, we show the incentive security properties of our full mechanisms, with consideration of collusions. 

%In Section~\ref{subsec:threat:model}, we model 4 types of attacks to the PoL mechanism. In the protocol defined in Section~\ref{subsec:prover:prot}, the training task is divided into $T$ stages. Even though it is a \emph{stochastic} gradient descent task, since the random seeds are given by the protocol, the training process of each stage is deterministic.

\subsection{Prover's Incentive Security}
\label{subsec:pis}

In the prover's training process, the prover is expected to save the model weights $W_t$ at each stage $t$, and post $c_{t}= hash(W_t)$. An honest prover should compute each $W_t$ from the result $W_{t-1}$ of the previous stage following the expected procedure.

For a possibly dishonest prover, in each stage $t$, she may compute $W_t$ from $W_{t-1}$ either honestly or dishonestly, or even does not compute a $W_t$ at all while forging a fake $c_t$. In our definition, even if $W_{t-1}$ may be dishonestly computed, as long as she follows the procedure and computes $W_t$ from $W_{t-1}$, we say that she trains stage $t$ ``honestly''; otherwise, if either $W_{t-1}$ or $W_t$ is nonexistent or invalid, or the prover does not follow the procedure when computing $W_t$ from $W_{t-1}$, we say that she trains the stage $t$ ``dishonestly''. Hence, we can naturally define the $\rho$ (as discussed in Section~\ref{subsec:inc:model}) as the fraction of stages trained honestly and say that the prover is honest if and only if $\rho=1$, i.e., she trains all stages honestly.

As introduced in Section~\ref{subsec:verifier:prot}, the verifier randomly chooses $\alpha$ stages among the $T$ stages to verify. For each chosen stage $t$, the verifier queries the prover for $(W_{t-1},W_{t})$ and verifies if $W_{t-1}$, $W_{t}$ match the hashes and $W_{t}$ is the result of honest computation from $W_{t-1}$. Since the prover needs to post hashes of weights before the verification, all the weights have to be finalized before the verification. Hence, the prover would pass the verification with a probability of $1$ if and only if all verified stages are trained honestly. In the full mechanism, if $\xi$ verified stages are not trained honestly, the prover passes the verification with a probability of $2^{-\xi}$.

In Section~\ref{subsec:threat:model} we discussed about 4 types of attacks. In retraining-based spoofing, the attacker aims to get the same PoL, while in the other 3 types of attacks, the attacker aims to get a different PoL. Due to the deterministic nature of our protocol, if the attacker aims to get a different PoL, she must train a subset of stages dishonestly, which is indeed classified as ``dishonest'' in our analysis\footnote{In the augmentation of Section~\ref{sec:verIS} there may exist different valid \emph{safe deviations} but they could not save any computational cost.}; for the retraining-based spoofing, since the attacker aims to get the same PoL, it can neither save any computational cost nor corrupt the model, so it only has interest in copyright protection and does not need to be considered for the motivation of out setting that aims to adopt PoL as a PoUW.

For prevention of the 3 types of attacks, under mild assumptions, we show that our mechanism is incentive-secure for small $\alpha$ compared to the number $T$ of stages and a moderately large $R_1$, as characterized as below:

\begin{itemize}
\item Even with no penalty $(\gamma=0)$, an $\alpha = O(\log T)$ is sufficient as long as the reward $R$ guarantees ``just slightly more than'' individual-rationality.
\item With moderate penalty $\gamma=\Theta(1)$, an $\alpha = O(1/\gamma) = O(1)$ and a reward $R$ guaranteeing IR are sufficient to guarantee $\gamma$-IR-BIS.
\item With $\eta \in \left[\frac{2\alpha}{T},\frac{1}{2}\right)$ and $R_1 \ge \frac{M}{T}\left(\frac{2}{\eta}+1\right)$, our full mechanism is guaranteed to be VIS.
\end{itemize}
% even with no penalty $(\gamma=0)$, an $\alpha = O(\log T)$ is sufficient for our mechanism to ensure basic incentive-security as long as the reward $R$ guarantees ``just slightly more than'' individual-rationality.
Formally, we have our main theorem on the prover side:

\begin{theorem}[Main Theorem]\label{thm:main}
    Assume $T\ge 2$, and denote $\beta = \frac{M}{R}$. If {the winning probability function $P(\cdot)$ is differentiable and its hazard rate is upper bounded by $\lambda$}, i.e.,
    \begin{equation}\label{eqn:hazard:rate}
        \frac{P'(\rho)}{P(\rho)} \in [-\lambda, 0], \forall \rho\in [0,1],
    \end{equation}
    {in which $P'(\cdot)$ is denoted as the derivative of $P(\cdot)$}; and in the verification protocol defined, a cheating stage has at least a $\kappa=\Theta(1)$ probability to be caught when verified, then the mechanisms defined as Algorithms~\ref{alg:p1}-\ref{alg:v1} and Algorithms~\ref{alg:p2}-\ref{alg:v2} are $0$-IR-BIS if
    \begin{equation}\label{eqn:ir}
        R > \frac{\int_{0}^1 P(\rho) d \rho \cdot M} {P(1)-(1-\kappa)^{\alpha}} ,
    \end{equation}
    \begin{equation}\label{eqn:bis2}
        \alpha \ge \max\left\{\frac{2(\lambda+\beta)}{\beta \kappa},\frac{2 \ln\frac{T}{\beta}}{\kappa}\right\},
    \end{equation}
    in which Eq.~\eqref{eqn:ir} exponentially converges to $R> \frac{\int_{0}^1 P(\rho) d \rho \cdot M} {P(1)}$, the sufficient and necessary condition for IR, when $\alpha$ is moderately large.
\end{theorem}
\iffalse
\begin{figure*}[bt]
  \centering
  \subfigure[$T=10^3$]{
    \begin{minipage}{0.321\linewidth}
    \centering
    \includegraphics[scale=0.41]{photos/wv/100_02_1000_.pdf}
  \end{minipage}
  }
\subfigure[$T=10^4$]{
    \begin{minipage}{0.321\linewidth}
    \centering
    \includegraphics[scale=0.41]{photos/wv/100_02.pdf}
  \end{minipage}
  }
    \subfigure[$T=10^5$]{
    \begin{minipage}{0.321\linewidth}
    \centering
    \includegraphics[scale=0.41]{photos/wv/100_02_100000_.pdf}
  \end{minipage}
  }
  \caption{Influence of $T$ to Verifier's Incentive in Constant Strategy}
  \label{fig:vi:T}
\end{figure*}
\fi
The proof is deferred to Appendix~\ref{app:main:proof}. From the main theorem, we see that for a fixed $P$, the number of required stages for verification is $O(\log T)$ for bounded $\lambda$ and $\kappa = \Theta(1)$, making the relative computational overhead as low as $O(\frac{\log T}{T})=O(\frac{k\log E}{E})$. %Noticing that the on-chain space consumption is $O(\frac{E}{k})$, we can make $k$ grow proportional to $\sqrt{E}$. In this way, the PoL protocol has a total cost scaling with $\sqrt{\lambda E}$, which is an $O(\sqrt{\lambda/E})$ fraction of the ML task.  

Furthermore, by inducing penalty $\gamma=\Theta(1)$, i.e. getting caught cheating leads to a penalty comparable to the block reward, we can lower the number of required stages to $O(1)$ and the relative computational overhead to $O(\frac{k}{E})$. Formally, we have:

\begin{theorem}\label{thm:penalty}
    For $\gamma>0$, with the same definition of $\beta,\lambda,\kappa$ as in Theorem~\ref{thm:main}, %and in the verification protocol defined in Section~\ref{subsec:verifier:prot}, a cheating stage has at least a $\kappa=\Theta(1)$ probability to be caught when verified, 
    the mechanisms defined as Algorithms~\ref{alg:p1}-\ref{alg:v1} and Algorithms~\ref{alg:p2}-\ref{alg:v2} are $\gamma$-IR-BIS if
    \begin{equation}\label{eqn:ir2}
        R > \frac{\int_{0}^1 P(\rho) d \rho \cdot M} {P(1)} ,
    \end{equation}
    \begin{equation}\label{eqn:bis3}
        \alpha > \max\left\{\frac{\beta}{\gamma \kappa}, \frac{\lambda}{\kappa}\right\}.
    \end{equation}
\end{theorem}

The proof of Theorem~\ref{thm:penalty} is deferred to Appendix~\ref{app:proof:penalty}.

\subsection{Prover's Collusion-Proofness}

\textbf{Values of $\kappa$.} In the basic mechanism that assumes the honesty of the verifier, $\kappa=1$. In the full mechanism with $n=5, m=3$, if all verifiers are honest, as the dishonest prover can commit a dishonest stage as a flag, and each verifier will have a $\frac{1}{2}$ probability to vote on a different flag as her commitment, we have
$\kappa = \sum_{i=0}^{2} \binom{5}{i}=\frac{1}{2}.$

\textbf{$\kappa$ in the presence of collusions.} We can also see that $\kappa=\Theta(1)$ even if the prover can collude with at most $2$ of the verifiers. We can compute that when $1$ or $2$ verifiers are in collusion with the prover, the $\kappa$ will decrease to
$\kappa_{1,5}=\sum_{i=0}^{1}\binom{4}{i}=\frac{5}{16}$
and
$\kappa_{2,5}=\sum_{i=0}^{0}\binom{3}{i}=\frac{1}{8}$
respectively, which means that we need to verify $60\%$ and $300\%$ more stages to prevent the prover from colluding with $1$ and $2$ verifiers, but the design still works as long as the voting scheme can manage to catch a dishonestly train stage with a constant probability.

\subsection{Verifier's Incentive Security \added{and Collusion-Proofness}}
\label{subsec:vis}

On the other hand, we show the verifier incentive-security property of our mechanism, which, combined with the basic incentive-security properties of our mechanism, guarantees a Nash equilibrium in which both parties behave honestly. \added{As the verifier's rewards are given stage-wise, in this part, we only consider one stage in the discussion of verifiers' incentives.}

While the canonical \emph{Nash equilibrium} is motivated to ensure that no player can benefit from individual deviation, in real-world decentralized systems, there may be collusions in which a subset of participants may manipulate their actions together in order to increase their total utility, like in the discussion of \cite{chung2023foundations}. To model the collusion, we define the term of $(N_p,N_v)$-collusion %and $(N_p,N_v)$-SCP 
as:

\begin{definition}[$(N_p,N_v)$-collusion]
    In an $(N_p,N_v)$-collusion, at most $N_p$ provers and $N_v$ verifiers may collude with each other, while other parties are honest.
\end{definition}

%\begin{definition}[$(N_p,N_v)$-SCP (side contract proof)]
    %A mechanism is $(N_p,N_v)$-SCP if and only if an $(N_p,N_v)$-collusion cannot benefit from dishonest actions.
%\end{definition}

In this setting, we can observe that when the verifiers verify an honest stage, all honest verifier will report the same type as the ground truth. Hence, as long as no more then $n-m$ verifiers are dishonest, they will reach a correct consensus per the protocol in Section~\ref{subsection:multi:verifier:mechanism}. Hence, we can deduce that:

\begin{proposition}
    In a $(0,n-m)$-collusion (implying the prover's honesty), all honest verifiers will be \acc ed; a dishonest verifier, whether lazy or malicious, will be \acc ed with the same probability as she reports the correct answer.
\end{proposition}

Hence, as long as the prover is honest and the verifier pool is not severely corrupted, the verifiers will gain positive expected utility when honest, and negative expected utility when dishonest. Formally, we have:

\begin{lemma}
    In a $(0,n-m)$-collusion, an honest verifier gets an expected utility of $\eta R_1-(1+\frac{\eta}{2})\frac{M}{T}$, and a dishonest verifier gets an expected utility of at most $\max\{-2\eta R_1,-(2-\frac{\eta}{2}) R_1\}<0$. \footnote{We ignore the effect of ``sampling without replacement'' that already verified stages will affect the flag rate of remaining stages. Actually, when $T$ is large, $\alpha=O(\log T)$, and the number of flags is $\eta T=\Theta(T)$, this effect is practically negligible. For strict rigor, we can replace the ``$\eta$'' with ``$(1\pm O(\frac{\log T}{T}))\eta$'' in consideration of this aspect.}
    \label{lem:verifier:util}
\end{lemma}

The proof of Lemma~\ref{lem:verifier:util} is deferred to Appendix~\ref{app:verifier:util}. Nevertheless, in practice, we need to consider the scenario in which a small fraction $\eps$ of the provers are dishonest. In this case, we consider the worst case for verifiers' incentives: according to Table~\ref{table:veri:reward}, the maximum reward a verifier can get is $R_1$ in any case, and the minimum reward is $-2R_1$. Hence, we observe that:

\begin{observation}
    When the prover is dishonest, an honest verifier receives a reward of at least $-2R_1$, whereas a dishonest verifier receives at most $R_1$. 
    \label{obs:dishonest:prover}
\end{observation}

\begin{small}
\begin{table}[tb]
%\vspace{-1em}
\centering
\begin{tabular}{|c|c|c|}
\hline
& Prover is Honest &  Prover is Dishonest \\ \hline
Be Honest & $\eta R_1-(1+\frac{\eta}{2})\frac{M}{T}$ & $\ge-2R_1$    \\ \hline
Be Dishonest  & $\le \max\{-2\eta R_1,-(2-\frac{\eta}{2}) R_1\}$ & $\le R_1$   \\ \hline
\end{tabular}
\caption{Verifier's Utility}\label{table:veri:util}
\end{table}
\end{small}

From Lemma~\ref{lem:verifier:util} and Observation~\ref{obs:dishonest:prover}, we can show the verifiers' utility in Table~\ref{table:veri:util}.
Therefore, intuitively, as long as $\eps$ (the fraction of dishonest provers) remains small, the influence of dishonest provers would not significantly affect the verifiers' incentive structures, and the verifiers will be robustly incentivized to verify honestly, in turn ensuring the prover's incentive security guarantees in Section~\ref{subsec:pis}.

\begin{theorem}\label{thm:vis}
If $\frac{M}{T}=cR_1$ for $c\in(0,\frac{4}{7})$, then for any $\eta\in(\frac{2c}{2-c},\frac{4}{5})$, 
our full mechanism defined as Algorithms~\ref{alg:full}-\ref{alg:v2} is $\eps$-VIS (as in Definition~\ref{def:vis}) even in the presence of at most $n-m$ colluding verifiers, in which
 \begin{equation}
    \eps = \frac{\eta-(1+\frac{\eta}{2})c}{3}.
 \end{equation}
\end{theorem}
The proof of Theorem~\ref{thm:vis} is deferred to Appendix~\ref{app:proof:vis}. Notice that a higher flag rate $\eta$ ensures better VIS properties in a higher fraction of dishonest provers, but it also incurs more verification overhead, which is a trade-off between robustness and efficiency. Furthermore, the constraint of $c<\frac{4}{7}$ requires the verification reward to be more than $1.75$ times of the verification cost, indicating that a necessary surplus for the verification task is essential to incentivize the verifiers to actively participate in the task, as a \emph{robustness margin} discussed in \cite{zhao2024takes}. Furthermore, we only incur a $2R_1$ penalty instead of high penalties as \cite{zhang2024proof}, which benefits robustness according to the \emph{compactness criteria} in \cite{zhao2024takes}. 

\section{Experimental Demonstration}
\label{sec:exp}

In this section, we perform experiments to evaluate the practical performance on our proposed PoL mechanism. In our experiments, we test our mechanism with the CIFAR (CNN) and MNIST (MLP) training tasks on a computer with  NVIDIA GeForce RTX 4090 and 24GB memory. Each task contains $T=E=1000$ stages with each stage containing one epoch, and for robustness of the system, each task is independently verified by $n=5$ verifiers. We set parameters $\eta=0.2$ ($20\%$ stages flagged), $\gamma=0$ (no penalty), $\beta=\frac{1}{2}$ (the reward is $2$ times the computational cost), and $\alpha \in \{1,2,5,8,10,50\}$ as the number of stages each verifier checks. 

\subsection{Experimental Results}

In the experiments, we perform the following groups of tests with different types of attacks as shown in Table~\ref{table:attacks}. Among these attacks, only the partial spoofing attack shows non-zero success rates, as other attacks invalidate the output of every stage and will be detected even if only one stage is checked. We notice that the attacks of \citep{zhang2022adversarial,fang2023proof} essentially modify the training process to exploit the error tolerance in the work of \citet{jia2021proof} and lie in the scope of \emph{distillation-based spoofing attack}, and hence are effectively prevented by our mechanism.

\begin{small}
\begin{table}[htb]
\vspace{-1em}
\centering
\begin{tabular}{|p{0.03\linewidth}|p{0.65\linewidth}|p{0.2\linewidth}|}
\hline
\# & Attack Type & Success Rate\\ 
\hline
$0$ & Honest: No cheating or attack. & $1$ \\
\hline
$1$ & Known-model Attack: The attacker submits a pretrained
model obtained from external sources. & $0$
\\ \hline
$2$ & Model-stealing Attack: The attacker submits a model
trained by others who received the same training task. & $0$
\\ \hline
$3$ & Stochastic Spoofing Attack: The attacker randomly generates formatmatched
results as the certificate.& $0$
\\ \hline
$4$ & Structurally Correct Spoofing Attack: The attacker
mimicks the format of a PoL,
randomly updating the model's weight without doing the
actual training.& $0$
\\ \hline
$5$ & Distillation-based Spoofing Attack: The attacker
modifies some parameters or the training process. \textbf{Attacks of \citep{zhang2022adversarial,fang2023proof} lie in this scope.} & $0$
\\ \hline
$6$ & Partial Spoofing Attack: The attacker trains partial of the stages honestly and partial dishonestly.& Depending on parameters.
\\ \hline

\end{tabular}
\caption{Types of attacks in the experiments.}
\label{table:attacks}
\vspace{-2.5em}
\end{table}

\end{small}

\textbf{Provers' rewards.} In Figures~\ref{fig:exp}, we show the experimental results for training CIFAR and MNIST datasets with different $\alpha$, in which the Proportional Rule is used for prover's rewards and the reward ratio refers to the expected reward from the system compared to honest training. From the plots we show that the system can detect almost all partial spoofs with $\alpha=50$, i.e. each verifier verifies $5\%$ of all stages. For smaller $\alpha$, the expected reward of a spoof increases with higher honesty ratios and decreases with larger $\alpha$'s. %Notice that the Strict-Proportional Rule results in lower rewards for dishonest provers, so incentive properties holding for the Proportional Rule also hold for the Strict-Proportional Rule naturally.

\begin{figure}[htbp]

\centering
\vspace{-0.7em}
\hspace{-1.5em}
\includegraphics[width=0.51\linewidth]{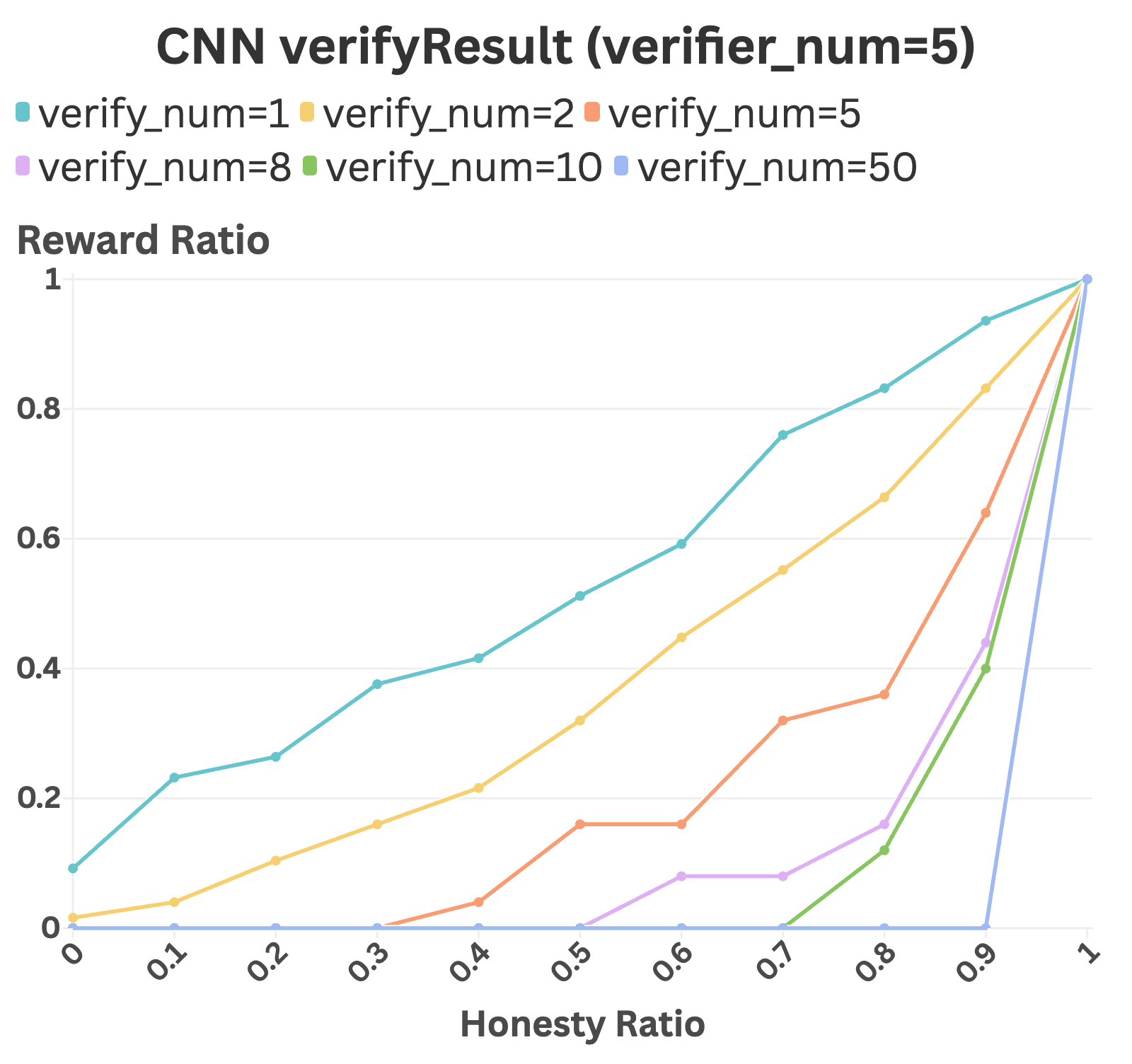}
\hspace{-0.6em}
\includegraphics[width=0.51\linewidth]{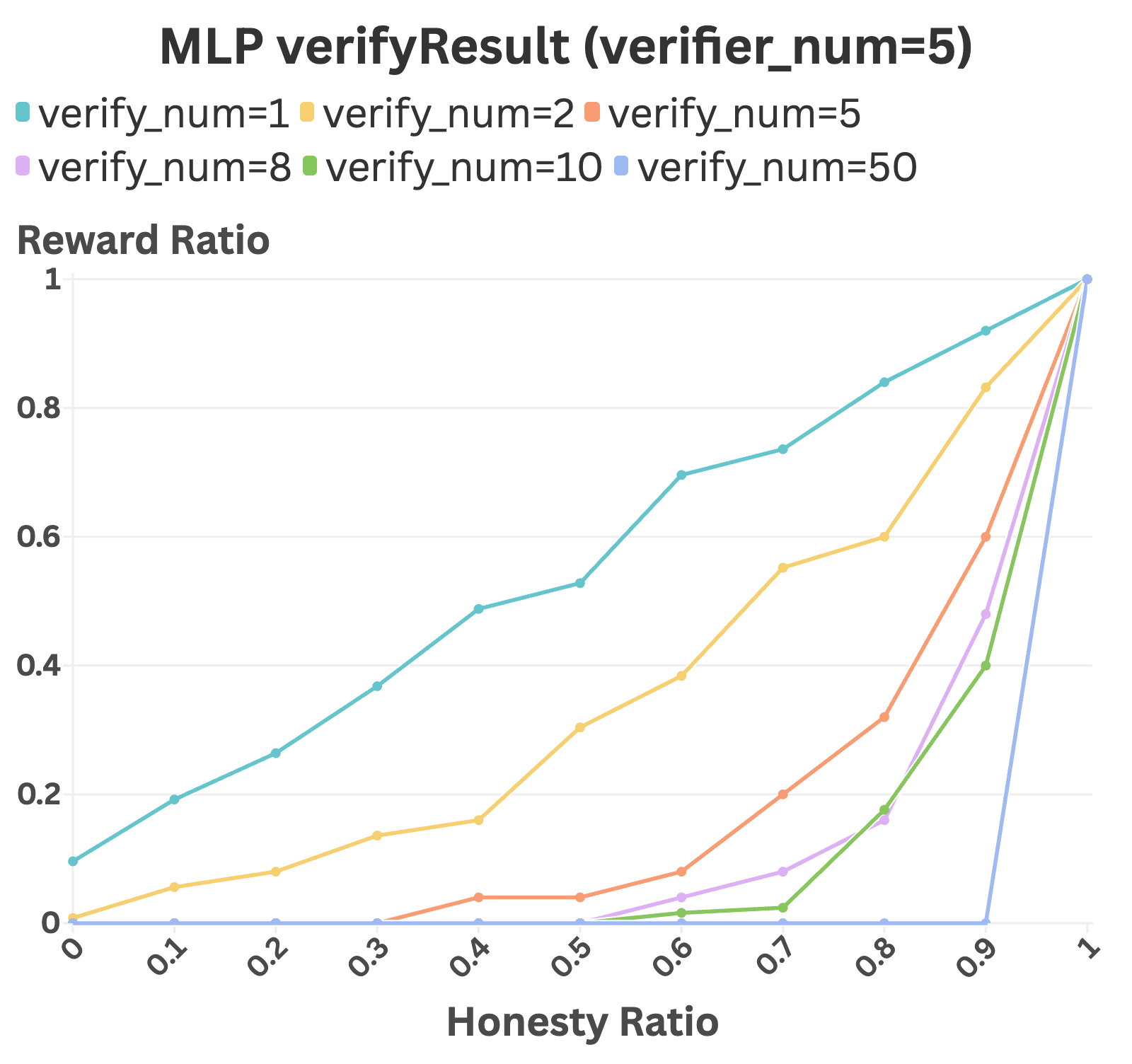}
\hspace{-1.5em}
\vspace{-1em}
\caption{Experimental Results.}
\label{fig:exp}
\vspace{-1.5em}
\end{figure}

\begin{figure}[htbp]

\centering
\hspace{-1.5em}
\includegraphics[width=0.51\linewidth]{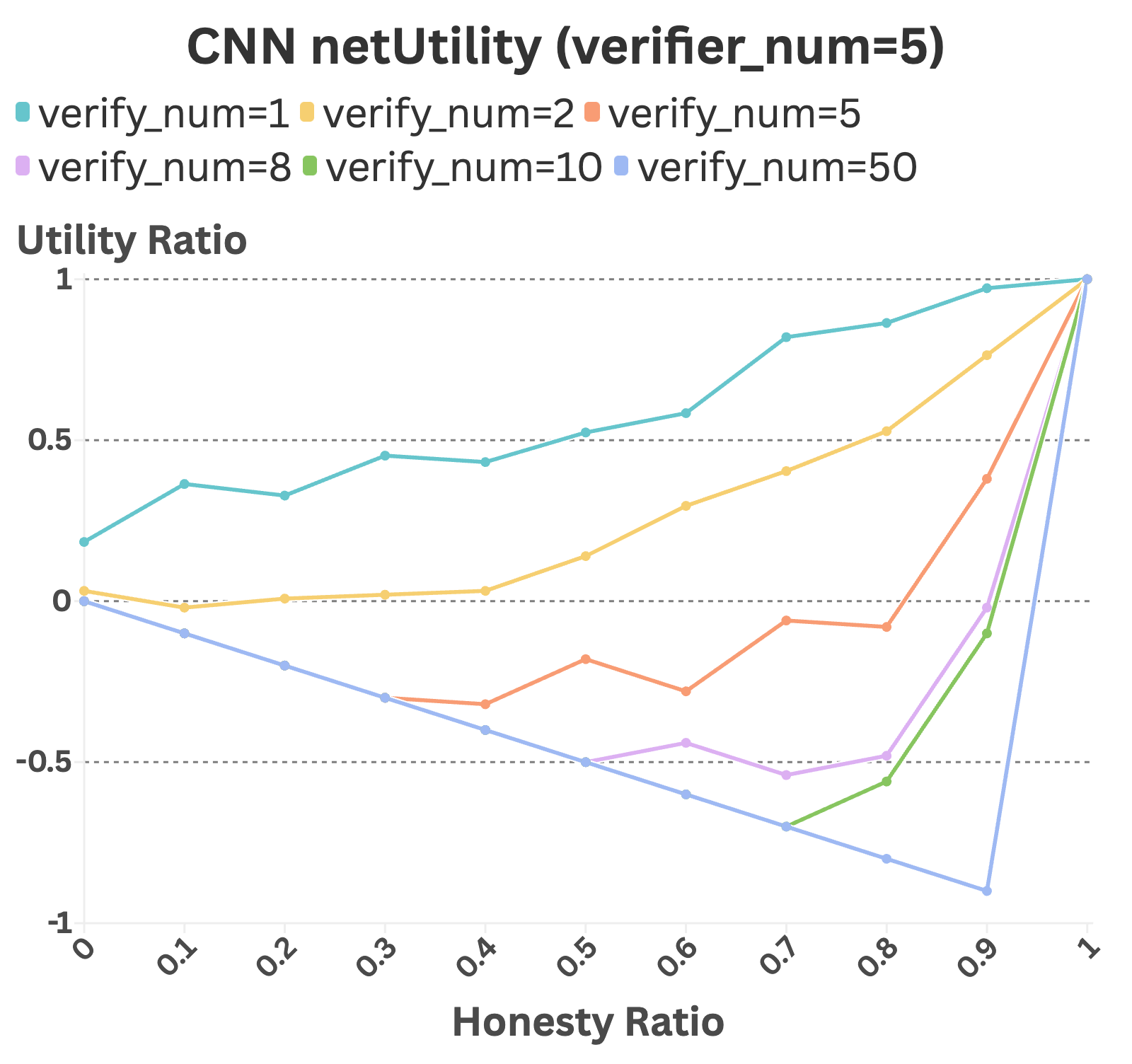}
\hspace{-0.6em}
\includegraphics[width=0.51\linewidth]{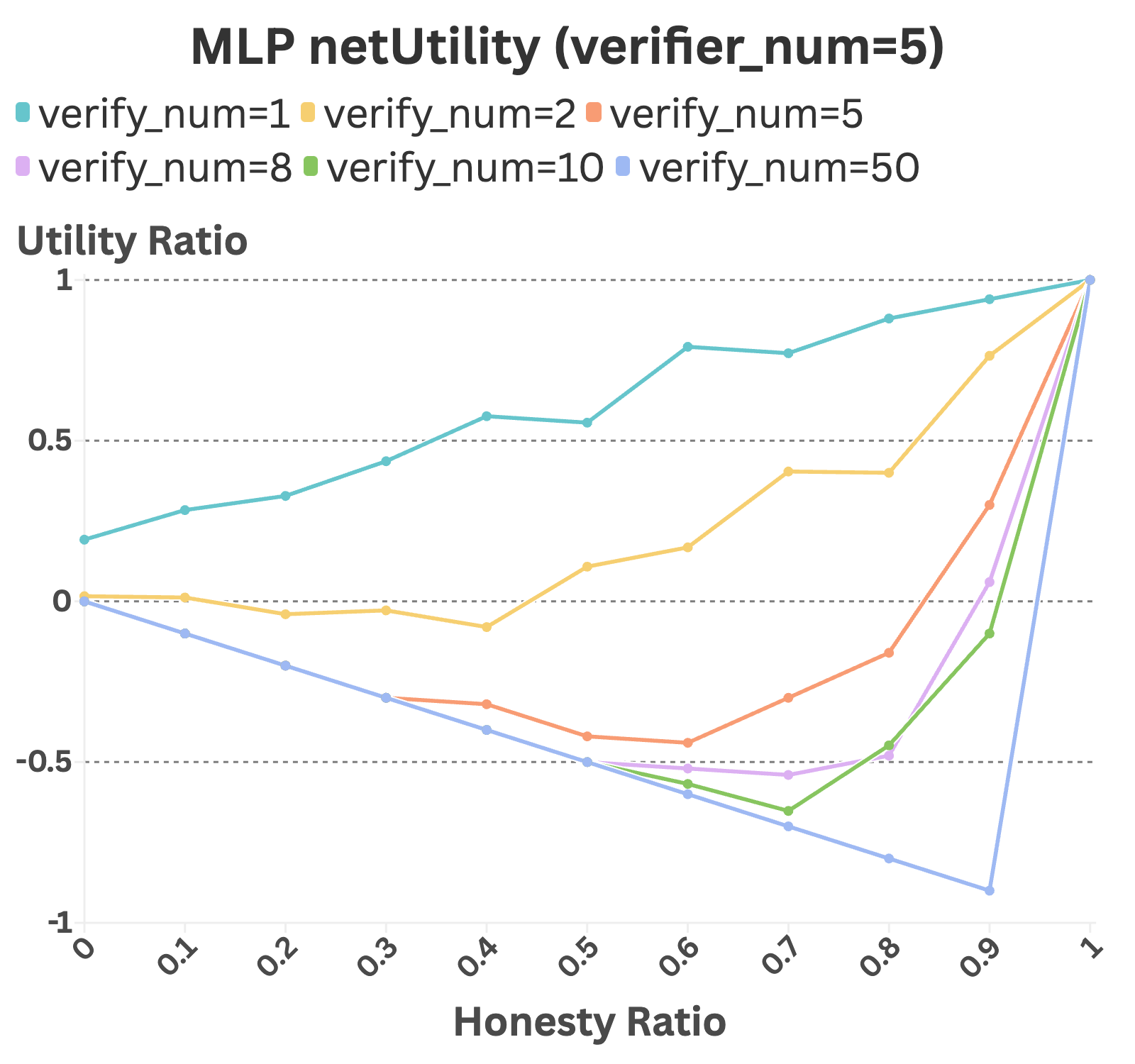}
\hspace{-1.5em}
\vspace{-1em}
\caption{Prover Net Utilities.}
\label{fig:exp:util}
\vspace{-1em}
\end{figure}

Furthermore, in Figure~\ref{fig:exp:util} we show the incentive properties of our mechanism for the tasks. The ``Utility Ratio'' refers to the net utility (reward minus computational cost) compared honest training. From the results, we see that when there is no mining competition, training the model honestly yields the maximum utility for the prover even for $\alpha=1$, i.e., the mechanism is incentive-secure. Furthermore, we see that for $\alpha\ge 10$, the prover gets negative utility unless at least $90\%$ of the stages are honestly trained, showing the sharpness of our incentive guarantee even for small $\alpha$'s.

%Since the experimental evaluation in the scenario with mining competition is complicated with real training tasks, particularly for the estimation of sunk costs when losing the competition, we refer to Theorems~\ref{thm:main}-\ref{thm:vis} for theoretical guarantees and leave real-world experiment for future empirical study.

%\textbf{Verifiers' rewards.} In the notion of Nash equilibria, we assume the honesty of the prover and all other verifiers. When we consider the case of $\alpha=50$ in which there is an overwhelming probability that all other verifiers report the ground truth (as shown in previous experiments), the proof is accepted and the verifier's reward and utility are simply proportional to the honestly verified stages. Hence, the verifier is indeed incentivized to honestly verify all $\alpha$ stages. In Appendix~\ref{app:vreward} we show the detailed experimental results and also demonstrate the necessity of the CTF protocol for the assurance of verifiers' incentive guarantees empirically. 

\textbf{Computational overheads.} In Table~\ref{table:overhead}, we show the average running time for training and verification, in which $\alpha=50$ epochs are verified among a total of $E=T=1000$. We can see that for each verifier, verifying an honest proof takes $6.1\%$ of the training time, slightly higher than $\frac{\alpha}{T}=5.0\%$ as the flag test in Algorithm~\ref{alg:v2} takes additional computation. Since we expect that most of the proofs are honest, our mechanism indeed achieves low computational overheads. %We also notice that the verification of dishonest proofs takes up to roughly $60\%$ more time as all dishonest stages trigger the flag test. A possible improvement for the incentive mechanism is to increase the verifier reward for rejected proofs to compensate for the additional computational costs.

\textbf{Communication overheads.} The communication overheads are shown in Table~\ref{tbl:comm}. We can see that the communication overheads are worse than computational overheads because full parameters need to be transmitted for verification, but still within a small fraction ($<20\%$) of all data generated during the training process. To further optimize the communication overheads, low-rank training techniques (e.g., GaLore \citep{zhao2024galore,zhang2024q}) can be adopted to optimize the overall I/O overheads for training tasks.

\begin{small}
\begin{table}[tb]
\centering
\begin{tabular}{|c|c|c|c|}
\hline
Honest Ratio & Training (s) & Verification (s) & Overhead/Verifier (\%)
\\ \hline
$0.0$ & $169.5$ & $282.3$ & $166.5$
\\ \hline
$0.1$ & $500.5$ & $274.5$ & $54.8$
\\ \hline
$0.2$ & $775.5$ & $264.4$ & $34.1$
\\ \hline
$0.3$ & $984.0$ & $253.1$ & $25.7$
\\ \hline
$0.4$ & $1235.5$ & $243.5$ & $19.7$
\\ \hline
$0.5$ & $1521.0$ & $228.6$ & $15.0$
\\ \hline
$0.6$ & $1717.5$ & $222.3$ & $12.9$
\\ \hline
$0.7$ & $2027.5$ & $213.0$ & $10.5$
\\ \hline
$0.8$ & $2356.5$ & $199.5$ & $8.5$
\\ \hline
$0.9$ & $2642.5$ & $185.8$ & $7.0$
\\ \hline
$1.0$ & $2782.0$ & $171.0$ & $6.1$
\\ \hline

\end{tabular}
\caption{Computational Overhead Analysis. ($\alpha=50$)}
\vspace{-2em}
\label{table:overhead}

\end{table}

\end{small}

\begin{small}
\begin{table}[tb]
%\vspace{-0.5em}
\centering
\begin{tabular}{|c|c|c|}
\hline
 & MNIST & CIFAR 
\\ \hline
Model Size (MB) & $52.41$ & $162.60$  
\\ \hline
Data Generated in Training (MB) &  $2369$ & $3595$  
\\ \hline
Transmission/Verifier (MB), $\alpha=10$ & $147$ &  $382$
\\ \hline
Overhead/Verifier ($\%$), $\alpha=10$ & $6.2$ & $10.6$ 
\\ \hline
Transmission/Verifier (MB), $\alpha=50$ & $333$ &  $658$
\\ \hline
Overhead/Verifier ($\%$), $\alpha=50$ & $14.0$ & $18.3$ 
\\ \hline
\end{tabular}
\caption{Communication Overhead Analysis.}

\label{tbl:comm}
\end{table}
\end{small}
\section{Discussion}

In this paper, we develop an incentive-secure PoL mechanism with provable incentive security, efficiency and controllable difficulty that bypasses existing hardness results, and also tackles the Verifier's Dilemma via a capture-the-flag protocol that encourages honest verification, while improving the relative computational overhead from $\Theta(1)$ in \citep{jia2021proof} to $O(\frac{\log E}{E})$ or $O(\frac{1}{E})$, and improving the communication complexity from $\Theta(E|\W|)$ in \citep{jia2021proof} to $O(E+|\W|\log E)$ or $O(E+|\W|)$, depending on different settings. On a high level, this paper not only provides an approach toward a secure and sustainable PoUW puzzle, but also works as a novel design for decentralized AI platforms.

%A possible issue is that while we assume the prover inserts flags and the verifier verifies randomly, a ``distribution drift'' may occur when in the long run all provers prefer to insert flags in a specific non-uniform distribution and verifiers also verify non-uniformly according to the pattern for optimal utility when \emph{the randomness is not enforced}. To resolve this issue, we may require that although we allow the prover/verifier to decide on the flag/verification positions with their own secrets, they must randomly draw the positions via a designated salted PRG with their committed seeds, so that they cannot find a seed that generates a statistically unbalanced distribution without excessive computational burden. 

While our mechanism can significantly improve the communication complexity compared to previous work, if the communication is implemented on-chain, it only works for relatively small models. To enable larger models compared to block spaces, IPFS \citep{benet2014ipfs} or layer-2 techniques \cite{sguanci2021layer} can be used for cheaper storage.

In real-world applications where the trained model may have exogenous interests, the prover may gain additional utility from training an incorrect model. In this scenario, our mechanism can be augmented with a family of anomaly detection techniques for deep learning \cite{paudice2018detection} and ensure that corrupting a small number of epochs would not significantly corrupt the output model.
%; if the attacker spoofs a large number of epochs, our mechanism can inherently detect the attacks with overwhelming probability.
%An intuitive approach is to leverage the Johnson–Lindenstrauss lemma \cite{venkatasubramanian2011johnson} to randomly project the weights into low dimensional spaces while keeping the distance information as additional (small-size) certificates in PoL, and utilize the smoothness of gradients to prevent the prover to deviate significantly in any stage. 
%Since our PoL protocol can detect dishonest training with high probability unless the number of dishonest stages is extremely small, the anomaly detection can effectively prevent (even irrational) adversarial attacks that might output a substantially incorrect model. 
We defer high-level discussions to Appendix~\ref{app:anomaly} and leave the detailed study for future research.

%Additionally, while this work is focused on the proof and verification of a specific ML task, we are aware that mining competition may still lead to energy waste. Via future studies on techniques of distributed computing and task allocation, the energy efficiency of our mechanism may be further improved.
\begin{acks}
Zishuo Zhao would like to appreciate Yongzheng Jia and Hongyin Chen for discussion on general topics of PoUW, Shuran Zheng and Yuqing Kong for discussion on incentive design and possible future directions. Zishuo Zhao is also grateful for the encouragement of all the individuals mentioned, which has been instrumental in maintaining his poise while exploring these novel and challenging fields. 
\end{acks}

\ifdefined\Anonymous
\clearpage
\fi

\bibliographystyle{ACM-Reference-Format}
\bibliography{references}

@article{chen2022crowdsourcing,
  title={Crowdsourcing Work as Mining: A Decentralized Computation and Storage Paradigm},
  author={Chen, Canhui and Cheng, Zerui and Qu, Shutong and Fang, Zhixuan},
  journal={arXiv preprint arXiv:2211.06669},
  year={2022}
}

@article{zkml,
  title={An Efficient and Extensible Zero-knowledge Proof Framework for Neural Networks},
  author={Lu, Tao and Wang, Haoyu and Qu, Wenjie and Wang, Zonghui and He, Jinye and Tao, Tianyang and Chen, Wenzhi and Zhang, Jiaheng},
  journal={Cryptology ePrint Archive},
  year={2024}
}

@article{conway2024opml,
  title={opML: Optimistic Machine Learning on Blockchain},
  author={Conway, KD and So, Cathie and Yu, Xiaohang and Wong, Kartin},
  journal={arXiv preprint arXiv:2401.17555},
  year={2024}
}

@article{cao2025sedulity,
  title={SEDULity: A Proof-of-Learning Framework for Distributed and Secure Blockchains with Efficient Useful Work},
  author={Cao, Weihang and Doger, Mustafa and Ulukus, Sennur},
  journal={arXiv preprint arXiv:2512.13666},
  year={2025}
}

@inproceedings{li2023wrong,
  title={Am I wrong, or is the autograder wrong? Effects of AI grading mistakes on learning},
  author={Li, Tiffany Wenting and Hsu, Silas and Fowler, Max and Zhang, Zhilin and Zilles, Craig and Karahalios, Karrie},
  booktitle={Proceedings of the 2023 ACM Conference on International Computing Education Research-Volume 1},
  pages={159--176},
  year={2023}
}

@article{bhatore2020machine,
  title={Machine learning techniques for credit risk evaluation: a systematic literature review},
  author={Bhatore, Siddharth and Mohan, Lalit and Reddy, Y Raghu},
  journal={Journal of Banking and Financial Technology},
  volume={4},
  number={1},
  pages={111--138},
  year={2020},
  publisher={Springer}
}

@article{lyu2024keeping,
  title={Keeping llms aligned after fine-tuning: The crucial role of prompt templates},
  author={Lyu, Kaifeng and Zhao, Haoyu and Gu, Xinran and Yu, Dingli and Goyal, Anirudh and Arora, Sanjeev},
  journal={arXiv preprint arXiv:2402.18540},
  year={2024}
}

@article{ji2023ai,
  title={Ai alignment: A comprehensive survey},
  author={Ji, Jiaming and Qiu, Tianyi and Chen, Boyuan and Zhang, Borong and Lou, Hantao and Wang, Kaile and Duan, Yawen and He, Zhonghao and Zhou, Jiayi and Zhang, Zhaowei and others},
  journal={arXiv preprint arXiv:2310.19852},
  year={2023}
}

@book{hendrycks2024introduction,
  title={Introduction to AI safety, ethics and society},
  author={Hendrycks, Dan},
  year={2024},
  publisher={Dan Hendrycks}
}

@article{ren2024safetywashing,
  title={Safetywashing: Do AI Safety Benchmarks Actually Measure Safety Progress?},
  author={Ren, Richard and Basart, Steven and Khoja, Adam and Gatti, Alice and Phan, Long and Yin, Xuwang and Mazeika, Mantas and Pan, Alexander and Mukobi, Gabriel and Kim, Ryan H and others},
  journal={arXiv preprint arXiv:2407.21792},
  year={2024}
}

@article{zhao2024galore,
  title={Galore: Memory-efficient llm training by gradient low-rank projection},
  author={Zhao, Jiawei and Zhang, Zhenyu and Chen, Beidi and Wang, Zhangyang and Anandkumar, Anima and Tian, Yuandong},
  journal={arXiv preprint arXiv:2403.03507},
  year={2024}
}

@article{zhang2024q,
  title={Q-galore: Quantized galore with int4 projection and layer-adaptive low-rank gradients},
  author={Zhang, Zhenyu and Jaiswal, Ajay and Yin, Lu and Liu, Shiwei and Zhao, Jiawei and Tian, Yuandong and Wang, Zhangyang},
  journal={arXiv preprint arXiv:2407.08296},
  year={2024}
}

@article{bengio2024international,
  title={International Scientific Report on the Safety of Advanced AI (Interim Report)},
  author={Bengio, Yoshua and Mindermann, S{\"o}ren and Privitera, Daniel and Besiroglu, Tamay and Bommasani, Rishi and Casper, Stephen and Choi, Yejin and Goldfarb, Danielle and Heidari, Hoda and Khalatbari, Leila and others},
  journal={arXiv preprint arXiv:2412.05282},
  year={2024}
}

@inproceedings{chung2023foundations,
  title={Foundations of transaction fee mechanism design},
  author={Chung, Hao and Shi, Elaine},
  booktitle={Proceedings of the 2023 Annual ACM-SIAM Symposium on Discrete Algorithms (SODA)},
  pages={3856--3899},
  year={2023},
  organization={SIAM}
}

@article{zhao2024takes,
  title={It Takes Two: A Peer-Prediction Solution for Blockchain Verifier's Dilemma},
  author={Zhao, Zishuo and Chen, Xi and Zhou, Yuan},
  journal={arXiv preprint arXiv:2406.01794},
  year={2024}
}

@Article{BytedanceAttack,
  author =       {Reuters},
  title =        {ByteDance seeks \$1.1 mln damages from intern in AI breach case, report says},
  journaltitle = {Reuters},
  year =   {2024},
  note =      {November 28, 2024}
  }

@article{firt2023calibrating,
  title={Calibrating machine behavior: a challenge for AI alignment},
  author={Firt, Erez},
  journal={Ethics and Information Technology},
  volume={25},
  number={3},
  pages={42},
  year={2023},
  publisher={Springer}
}

@article{bengio2024managing,
  title={Managing extreme AI risks amid rapid progress},
  author={Bengio, Yoshua and Hinton, Geoffrey and Yao, Andrew and Song, Dawn and Abbeel, Pieter and Darrell, Trevor and Harari, Yuval Noah and Zhang, Ya-Qin and Xue, Lan and Shalev-Shwartz, Shai and others},
  journal={Science},
  volume={384},
  number={6698},
  pages={842--845},
  year={2024},
  publisher={American Association for the Advancement of Science}
}

@article{langenbucher2022ai,
  title={AI credit scoring and evaluation of creditworthiness--a test case for the EU proposal for an AI Act},
  author={Langenbucher, Katja},
  journal={how the challenges of today prepare the ground for tomorrow},
  pages={362},
  year={2022}
}

@article{tomic2022ai,
  title={An AI-based approach for grading students’ collaboration},
  author={Tomi{\'c}, Bojan B and Kijev{\v{c}}anin, Anisja D and {\v{S}}evarac, Zoran V and Jovanovi{\'c}, Jelena M},
  journal={IEEE Transactions on Learning Technologies},
  volume={16},
  number={3},
  pages={292--305},
  year={2022},
  publisher={IEEE}
}

@article{zhang2024proof,
  title={Proof of Sampling: A Nash Equilibrium-Secured Verification Protocol for Decentralized Systems},
  author={Zhang, Yue and Wang, Shouqiao and Liu, Xiaoyuan and Tan, Sijun and Popa, Raluca Ada and Moallemi, Ciamac C},
  journal={arXiv preprint arXiv:2405.00295},
  year={2024}
}

@article{riley2021current,
  title={The current status of cryptocurrency regulation in China and its effect around the world},
  author={Riley, John},
  journal={China and WTO Review},
  volume={7},
  number={1},
  pages={135--152},
  year={2021}
}

@InProceedings{10.1007/978-3-030-17653-2_23,
author="Ganesh, Chaya
and Orlandi, Claudio
and Tschudi, Daniel",
editor="Ishai, Yuval
and Rijmen, Vincent",
title="Proof-of-Stake Protocols for Privacy-Aware Blockchains",
booktitle="Advances in Cryptology -- EUROCRYPT 2019",
year="2019",
publisher="Springer International Publishing",
address="Cham",
pages="690--719",
isbn="978-3-030-17653-2"
}

@article{10.1093/rfs/hhaa075,
    author = {Saleh, Fahad},
    title = "{Blockchain without Waste: Proof-of-Stake}",
    journal = {The Review of Financial Studies},
    volume = {34},
    number = {3},
    pages = {1156-1190},
    year = {2020},
    month = {07},
    issn = {0893-9454},
    doi = {10.1093/rfs/hhaa075},
    url = {https://doi.org/10.1093/rfs/hhaa075},
    eprint = {https://academic.oup.com/rfs/article-pdf/34/3/1156/36264598/hhaa075.pdf},
}

@inproceedings{10.1145/3524458.3547248,
author = {Toulemonde, Ambre and Besson, Loic and Goubin, Louis and Patarin, Jacques},
title = {Useful work: a new protocol to ensure usefulness of PoW-based consensus for blockchain},
year = {2022},
isbn = {9781450392846},
publisher = {Association for Computing Machinery},
address = {New York, NY, USA},
url = {https://doi.org/10.1145/3524458.3547248},
doi = {10.1145/3524458.3547248},
booktitle = {Proceedings of the 2022 ACM Conference on Information Technology for Social Good},
pages = {308–314},
numpages = {7},
location = {Limassol, Cyprus},
series = {GoodIT '22}
}

@inproceedings{smuseva2022verifier,
  title={Verifier’s dilemma in ethereum blockchain: A quantitative analysis},
  author={Smuseva, Daria and Malakhov, Ivan and Marin, Andrea and van Moorsel, Aad and Rossi, Sabina},
  booktitle={International Conference on Quantitative Evaluation of Systems},
  pages={317--336},
  year={2022},
  organization={Springer}
}

@inproceedings{gervais2016security,
  title={On the security and performance of proof of work blockchains},
  author={Gervais, Arthur and Karame, Ghassan O and W{\"u}st, Karl and Glykantzis, Vasileios and Ritzdorf, Hubert and Capkun, Srdjan},
  booktitle={Proceedings of the 2016 ACM SIGSAC conference on computer and communications security},
  pages={3--16},
  year={2016}
}

@misc{sguanci2021layer,
      title={Layer 2 Blockchain Scaling: a Survey}, 
      author={Cosimo Sguanci and Roberto Spatafora and Andrea Mario Vergani},
      year={2021},
      eprint={2107.10881},
      archivePrefix={arXiv},
      primaryClass={cs.DC}
}

@article{benet2014ipfs,
  title={Ipfs-content addressed, versioned, p2p file system},
  author={Benet, Juan},
  journal={arXiv preprint arXiv:1407.3561},
  year={2014}
}

@inproceedings{10.1145/2976749.2978341,
author = {Gervais, Arthur and Karame, Ghassan O. and W\"{u}st, Karl and Glykantzis, Vasileios and Ritzdorf, Hubert and Capkun, Srdjan},
title = {On the Security and Performance of Proof of Work Blockchains},
year = {2016},
isbn = {9781450341394},
publisher = {Association for Computing Machinery},
address = {New York, NY, USA},
url = {https://doi.org/10.1145/2976749.2978341},
doi = {10.1145/2976749.2978341},
booktitle = {Proceedings of the 2016 ACM SIGSAC Conference on Computer and Communications Security},
pages = {3–16},
numpages = {14},
location = {Vienna, Austria},
series = {CCS '16}
}

@book{piketty2014capital,
  title={Capital in the twenty-first century},
  author={Piketty, Thomas},
  year={2014},
  publisher={Harvard University Press}
}

@inproceedings{kiayias2017ouroboros,
  title={Ouroboros: A provably secure proof-of-stake blockchain protocol},
  author={Kiayias, Aggelos and Russell, Alexander and David, Bernardo and Oliynykov, Roman},
  booktitle={Annual international cryptology conference},
  pages={357--388},
  year={2017},
  organization={Springer}
}

@inproceedings{jakobsson1999proofs,
  title={Proofs of work and bread pudding protocols},
  author={Jakobsson, Markus and Juels, Ari},
  booktitle={Secure Information Networks: Communications and Multimedia Security IFIP TC6/TC11 Joint Working Conference on Communications and Multimedia Security (CMS’99) September 20--21, 1999, Leuven, Belgium},
  pages={258--272},
  year={1999},
  organization={Springer}
}

@article{buterin2014next,
  title={A next-generation smart contract and decentralized application platform},
  author={Buterin, Vitalik and others},
  journal={white paper},
  volume={3},
  number={37},
  pages={2--1},
  year={2014}
}

@article{nakamoto2008bitcoin,
  title={Bitcoin: A peer-to-peer electronic cash system},
  author={Nakamoto, Satoshi},
  journal={Decentralized business review},
  year={2008}
}

@inproceedings{hoffmann2022challenges,
  title={Challenges of Proof-of-Useful-Work (PoUW)},
  author={Hoffmann, Felix},
  booktitle={2022 IEEE 1st Global Emerging Technology Blockchain Forum: Blockchain \& Beyond (iGETblockchain)},
  pages={1--5},
  year={2022},
  organization={IEEE}
}

@inproceedings{jia2021proof,
  title={Proof-of-learning: Definitions and practice},
  author={Jia, Hengrui and Yaghini, Mohammad and Choquette-Choo, Christopher A and Dullerud, Natalie and Thudi, Anvith and Chandrasekaran, Varun and Papernot, Nicolas},
  booktitle={2021 IEEE Symposium on Security and Privacy (SP)},
  pages={1039--1056},
  year={2021},
  organization={IEEE}
}

@inproceedings{zhang2022adversarial,
  title={“Adversarial Examples” for Proof-of-Learning},
  author={Zhang, Rui and Liu, Jian and Ding, Yuan and Wang, Zhibo and Wu, Qingbiao and Ren, Kui},
  booktitle={2022 IEEE Symposium on Security and Privacy (SP)},
  pages={1408--1422},
  year={2022},
  organization={IEEE}
}

@inproceedings{fang2023proof,
  title={Proof-of-Learning is Currently More Broken Than You Think},
  author={Fang, Congyu and Jia, Hengrui and Thudi, Anvith and Yaghini, Mohammad and Choquette-Choo, Christopher A and Dullerud, Natalie and Chandrasekaran, Varun and Papernot, Nicolas},
  booktitle={2023 IEEE 8th European Symposium on Security and Privacy (EuroS\&P)},
  pages={797--816},
  year={2023},
  organization={IEEE}
}

@inproceedings{fiz2020sluggish,
  title={Sluggish mining: Profiting from the verifier’s dilemma},
  author={Fiz Pontiveros, Beltr{\'a}n Borja and Ferreira Torres, Christof and State, Radu},
  booktitle={Financial Cryptography and Data Security: FC 2019 International Workshops, VOTING and WTSC, St. Kitts, St. Kitts and Nevis, February 18--22, 2019, Revised Selected Papers 23},
  pages={67--81},
  year={2020},
  organization={Springer}
}

@inproceedings{kiayias2020non,
  title={Non-interactive proofs of proof-of-work},
  author={Kiayias, Aggelos and Miller, Andrew and Zindros, Dionysis},
  booktitle={Financial Cryptography and Data Security: 24th International Conference, FC 2020, Kota Kinabalu, Malaysia, February 10--14, 2020 Revised Selected Papers 24},
  pages={505--522},
  year={2020},
  organization={Springer}
}

@inproceedings{bertino2021ai,
  title={AI for Security and Security for AI},
  author={Bertino, Elisa and Kantarcioglu, Murat and Akcora, Cuneyt Gurcan and Samtani, Sagar and Mittal, Sudip and Gupta, Maanak},
  booktitle={Proceedings of the Eleventh ACM Conference on Data and Application Security and Privacy},
  pages={333--334},
  year={2021}
}

@article{hu2021artificial,
  title={Artificial intelligence security: Threats and countermeasures},
  author={Hu, Yupeng and Kuang, Wenxin and Qin, Zheng and Li, Kenli and Zhang, Jiliang and Gao, Yansong and Li, Wenjia and Li, Keqin},
  journal={ACM Computing Surveys (CSUR)},
  volume={55},
  number={1},
  pages={1--36},
  year={2021},
  publisher={ACM New York, NY}
}

@article{khakurel2018rise,
  title={The rise of artificial intelligence under the lens of sustainability},
  author={Khakurel, Jayden and Penzenstadler, Birgit and Porras, Jari and Knutas, Antti and Zhang, Wenlu},
  journal={Technologies},
  volume={6},
  number={4},
  pages={100},
  year={2018},
  publisher={MDPI}
}

@article{vranken2017sustainability,
  title={Sustainability of bitcoin and blockchains},
  author={Vranken, Harald},
  journal={Current opinion in environmental sustainability},
  volume={28},
  pages={1--9},
  year={2017},
  publisher={Elsevier}
}

@inproceedings{ribeiro2015mlaas,
  title={Mlaas: Machine learning as a service},
  author={Ribeiro, Mauro and Grolinger, Katarina and Capretz, Miriam AM},
  booktitle={2015 IEEE 14th international conference on machine learning and applications (ICMLA)},
  pages={896--902},
  year={2015},
  organization={IEEE}
}

@article{stoll2019carbon,
  title={The carbon footprint of bitcoin},
  author={Stoll, Christian and Klaa{\ss}en, Lena and Gallersd{\"o}rfer, Ulrich},
  journal={Joule},
  volume={3},
  number={7},
  pages={1647--1661},
  year={2019},
  publisher={Elsevier}
}

@article{ibanez2023bitcoin,
  title={Bitcoin’s carbon footprint revisited: Proof of Work mining for renewable energy expansion},
  author={Iba{\~n}ez, Juan Ignacio and Freier, Alexander},
  journal={Challenges},
  volume={14},
  number={3},
  pages={35},
  year={2023},
  publisher={MDPI}
}

@article{paudice2018detection,
  title={Detection of adversarial training examples in poisoning attacks through anomaly detection},
  author={Paudice, Andrea and Mu{\~n}oz-Gonz{\'a}lez, Luis and Gyorgy, Andras and Lupu, Emil C},
  journal={arXiv preprint arXiv:1802.03041},
  year={2018}
}

@article{nassar2020blockchain,
  title={Blockchain for explainable and trustworthy artificial intelligence},
  author={Nassar, Mohamed and Salah, Khaled and ur Rehman, Muhammad Habib and Svetinovic, Davor},
  journal={Wiley Interdisciplinary Reviews: Data Mining and Knowledge Discovery},
  volume={10},
  number={1},
  pages={e1340},
  year={2020},
  publisher={Wiley Online Library}
}

@article{lee2019decentralized,
  title={A decentralized token economy: How blockchain and cryptocurrency can revolutionize business},
  author={Lee, Jei Young},
  journal={Business Horizons},
  volume={62},
  number={6},
  pages={773--784},
  year={2019},
  publisher={Elsevier}
}

@inproceedings{bagaria2022proof,
  title={Proof-of-stake longest chain protocols: Security vs predictability},
  author={Bagaria, Vivek and Dembo, Amir and Kannan, Sreeram and Oh, Sewoong and Tse, David and Viswanath, Pramod and Wang, Xuechao and Zeitouni, Ofer},
  booktitle={Proceedings of the 2022 ACM Workshop on Developments in Consensus},
  pages={29--42},
  year={2022}
}

@article{king2013primecoin,
  title={Primecoin: Cryptocurrency with prime number proof-of-work},
  author={King, Sunny},
  journal={July 7th},
  volume={1},
  number={6},
  year={2013}
}

@article{ileri2016coinami,
  title={Coinami: a cryptocurrency with DNA sequence alignment as proof-of-work},
  author={Ileri, Atalay M and Ozercan, Halil I and Gundogdu, Alper and Senol, Ahmet K and Ozkaya, M Yusuf and Alkan, Can},
  journal={arXiv preprint arXiv:1602.03031},
  year={2016}
}

@article{baldominos2019coinai,
  title={Coin. AI: A proof-of-useful-work scheme for blockchain-based distributed deep learning},
  author={Baldominos, Alejandro and Saez, Yago},
  journal={Entropy},
  volume={21},
  number={8},
  pages={723},
  year={2019},
  publisher={MDPI}
}

@book{goodhart1984problems,
  title={Problems of monetary management: the UK experience},
  author={Goodhart, Charles AE},
  year={1984},
  publisher={Springer}
}

@article{ball2017proofs,
  title={Proofs of useful work},
  author={Ball, Marshall and Rosen, Alon and Sabin, Manuel and Vasudevan, Prashant Nalini},
  journal={Cryptology ePrint Archive},
  year={2017}
}

@article{liu2021proof,
  title={Proof of Learning (PoLe): empowering neural network training with consensus building on blockchains},
  author={Liu, Yuan and Lan, Yixiao and Li, Boyang and Miao, Chunyan and Tian, Zhihong},
  journal={Computer Networks},
  volume={201},
  pages={108594},
  year={2021},
  publisher={Elsevier}
}

@article{reynouard2024bar,
  title={BAR Nash Equilibrium and Application to Blockchain Design},
  author={Reynouard, Maxime and Laraki, Rida and Gorelkina, Olga},
  journal={arXiv preprint arXiv:2401.16856},
  year={2024}
}

@incollection{teutsch2024scalable,
  title={A scalable verification solution for blockchains},
  author={Teutsch, Jason and Reitwie{\ss}ner, Christian},
  booktitle={ASPECTS OF COMPUTATION AND AUTOMATA THEORY WITH APPLICATIONS},
  pages={377--424},
  year={2024},
  publisher={World Scientific}
}

@inproceedings{luu2015demystifying,
  title={Demystifying incentives in the consensus computer},
  author={Luu, Loi and Teutsch, Jason and Kulkarni, Raghav and Saxena, Prateek},
  booktitle={Proceedings of the 22Nd acm sigsac conference on computer and communications security},
  pages={706--719},
  year={2015}
}

@inproceedings{zheng2020difficulty,
  title={Difficulty prediction for proof-of-work based blockchains},
  author={Zheng, Kaiwen and Zhang, Shulai and Ma, Xiaoli},
  booktitle={2020 IEEE 21st International Workshop on Signal Processing Advances in Wireless Communications (SPAWC)},
  pages={1--5},
  year={2020},
  organization={IEEE}
}

@inproceedings{bravo2019proof,
  title={Proof-of-learning: a blockchain consensus mechanism based on machine learning competitions},
  author={Bravo-Marquez, Felipe and Reeves, Steve and Ugarte, Martin},
  booktitle={2019 IEEE International Conference on Decentralized Applications and Infrastructures (DAPPCON)},
  pages={119--124},
  year={2019},
  organization={IEEE}
}

\clearpage

\appendix

\begin{center}
    \Large \textbf{Appendix}
\end{center}

\section{Computation of Prover's Sunk Cost $\mu(\rho)$ on Losing Competition}
\label{app:sunk:cost}

Define $P_-(t)$ as the probability that another prover would have finished the computation by the time the fixed prover computes a $t$ portion of the task. Then by definition, we have

\begin{equation}
    P_-(t) = 1 - P(t).
\end{equation}

Denote $X$ as the random variable of the portion the fixed prover has done to the task when another prover would submit the work, then $P_-(\cdot)$ is essentially the CDF of $X$, and the PDF of $X$ is $P_-'(\cdot)$.

Given that the fixed prover would stop computing when some other prover submits the task, we get that

\begin{align}
    \frac{\mu(\rho)}{M} &= \mathbb{E}[X | X < \rho] \\
    &= \frac{\mathbb{E}[X \cdot \mathbf{1}_{[ X < \rho]}]}{\Pr[X<\rho]} \\
    &= \frac{\int_0^{\rho} tP'(t) dt}{1-P(\rho)} \\
    &= \frac{\int_0^{\rho} \int_0^t P_-'(t) dx dt}{1-P(\rho)} \\
    &= \frac{\int_0^{\rho} \int_t^\rho P_-'(t) dt dx}{1-P(\rho)} \\
    &= \frac{\int_0^{\rho} (P_-(\rho)-P_-(x)) dx}{1-P(\rho)} \\
    &= \frac{\int_0^{\rho} (P(x) - P(\rho)) dx}{1-P(\rho)} \\
    &= \frac{\int_0^{\rho} P(x) dx - \rho P(\rho)}{1-P(\rho)}.
\end{align}

Therefore, 

\begin{equation}
\mu(\rho) = \frac{\int_0^{\rho} P(x) dx - \rho P(\rho)}{1-P(\rho)}M.
\end{equation}

\section{Ablation Analysis on Verifiers' Incentives}
\label{app:vreward}

\iffalse
We consider the case of $\alpha=50$ that the mechanism almost always makes the correct decision, as shown in Section~\ref{sec:exp}, and we set the expected verification reward to be $2$ times the verification cost of honest proofs. We can expect that there is an overwhelming probabilities that other players are honest. Hence, we assume that other $4$ of the $5$ verifiers are honest, and the proof is honest with probability $p_\textit{proof}\in[0,1]$ in increments of $0.2$; dishonest provers conduct partial spoof attacks with honest ratio $\rho=0.9$ (which is relatively hard to detect). Then, we run numerical simulations and plot the verifier's expected utility when she honestly verifies $\alpha'\in[0,50]$ stages in Figure~\ref{fig:vutil}. 

From Figure~\ref{fig:vutil} we see that for $p_\textit{proof}\ge 0.4$, the CTF protocol incentivizes the verifier to honestly verify all $\alpha=50$ stages via the flag rewards. For low $p_\textit{proof}$ (which is unlikely to occur due to prover-side incentive-security), the verifier is incentivized to verify fewer stages. The intuitive explanation is that verification rewards for rejected proofs are irrelevant to flags, and verifying $20$ to $30$ stages is already enough to detect the cheats with high probability.  
\fi

To empirically show the necessity of our CTF protocol, we perform numerical analysis as an ablation study. We consider the case of $\alpha=50$ that the mechanism almost always makes the correct decision, as shown in Section~\ref{sec:exp}, and we set the expected verification reward to be $2$ times the verification cost of honest proofs. We can expect that there is an overwhelming probabilities that other players are honest. Hence, we assume that other $4$ of the $5$ verifiers are honest, and the proof is honest with probability $p_\textit{proof}\in[0,1]$ in increments of $0.2$; dishonest provers conduct partial spoof attacks with honest ratio $\rho=0.9$ (which is relatively hard to detect). We plot the verifiers' utilities in Figure~\ref{fig:vutil2} when we use the basic mechanism (Algorithms~\ref{alg:p1}-\ref{alg:v1}) with verifiers' rewards given by simple majority vote. In the figure, we see that particularly for $p_\textit{proof}= 1$, the verifier would be incentivized to lazily accept the proof even if all other verifiers are honest, demonstrating the phenomenon of the Verifier's Dilemma. Hence, we show the practical effectiveness and necessity of our CTF protocol for the incentive guarantees on the verifier's side.

The intuitive explanation is that in the case that when there are few dishonest provers, there is no incentive of verifying the proof compared to lazily accepting it; even if a substantial fraction of dishonest provers, verifying $20$ to $30$ stages is already enough to detect the cheats with high probability, which is still ``\emph{partially lazy}'' behavior which undermines the trustworthiness and security of the PoL design.  

\iffalse

\begin{figure}[htb]
\centering

\includegraphics[width=0.95\linewidth]{photos/vutil.png}

\caption{Verifier's Utility}
\label{fig:vutil}
\vspace{-2em}
\end{figure}

\fi

\begin{figure}[htb]
\centering

\includegraphics[width=0.95\linewidth]{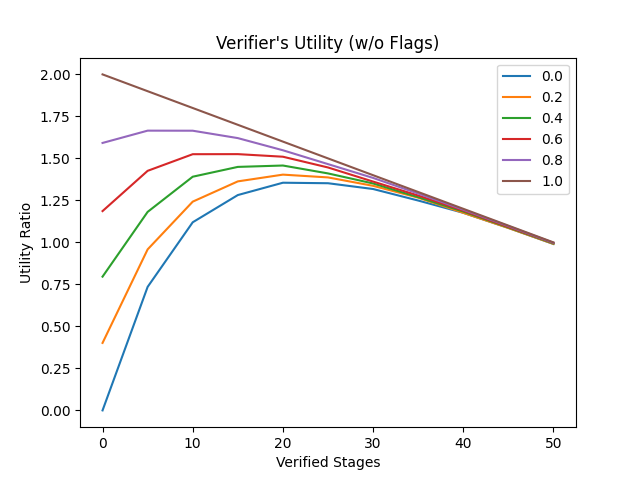}

\caption{Verifier's Utility without CTF Protocol}
\label{fig:vutil2}

\end{figure}

\section{{The Multi-Verifier Protocol}}
\label{subsection:multi:verifier:mechanism}

\begin{table*}[htb]
%\vspace{-1em}
\centering
\begin{tabular}{|c|c|c|c|c|c|}
\hline
& Quorum ``Normal'' & Quorum $F_1$ & Quorum $F_2$ & Quorum ``Invalid'' & No Quorum  \\ \hline
Report ``Normal'' & $0$ & $-2R_1$ & $-2R_1$ &$-2R_1$ &$-2R_1$   \\ \hline
Report $F_1$ & $-2R_1$ & $R_1$ & $-2R_1$ &$-2R_1$ &$-2R_1$ \\ \hline
Report $F_2$ & $-2R_1$ & $-2R_1$ & $R_1$ &$-2R_1$ &$-2R_1$  \\ \hline
Report  ``Invalid'' & $-2R_1$ & $-2R_1$ & $-2R_1$ &$R_1$ &$-2R_1$   \\ \hline
\end{tabular}

\caption{Verifier's Reward For Each Stage}\label{table:veri:reward}

\end{table*}

From the nature of the CTF protocol, we can observe that:

\begin{itemize}
\item If a stage is normal, then all honest verifiers who verify this stage will observe it as normal.
\item If a stage is honestly trained with flag $F_i$, then the prover and all honest verifiers who verify this stage will observe $F_i$.
\item If a stage is dishonestly trained, then an honest verifier who verifies this stage will observe $F_1$ or $F_2$ with probability $\frac{1}{2}$ each. 
\end{itemize}

Hence, we deduce that as long as the prover is honest, the prover and all honest verifiers that verify the same stage should reach an agreement. 

For simplicity, we assume that $n=5$ verifiers are involved in the same task to verify the same subset $\mathrm{t}_{ve}$ of stages, with a \emph{quorum} number of $m\ge 3$, and we design the voting process w.r.t. stage $t_i\in\t_{ve}$ as follows:

\begin{itemize}
\item If the prover's commitment agrees with at least $m$ verifiers' reports, \acc ~the prover and agreeing verifiers, and \rej~the other verifiers.
\item If at least $m$ verifiers' reports agree with each other but disagree with the prover, \acc~the agreeing verifiers, and \rej~the other verifiers and the prover.
\item Otherwise, \rej~the prover and every verifier.
\end{itemize}

To implement the voting scheme, we can let the verifiers join in the verification of the task freely until $n$ positions are filled, and they are supposed to verify {the same set of stages} via shared randomness, e.g., the hash value of the XOR sum of all the verifiers' IDs or submitted random numbers; after the verification, verifiers submit (broadcast) the \emph{commitments} of their reports until all $n$ commitments have been received for the task, to ensure that no verifier can see other verifiers' reports before submitting their own report; inactive verifiers are \emph{slashed} for failing to response. Then, the verifiers reveal their reports as the votes for the proof.

After the voting, we accept the prover if and only if he is \acc ed in every stage, rewarding him $R$; otherwise, reject the prover and reward him $-\gamma R$ (as modeled in Section~\ref{subsec:inc:model}). For the verifiers, we reward each verifier $R_1$ for each \emph{flag} \acc ed, and $-2R_1$ for each stage \rej ed. For clarity, we show the verifier's rewards for each verified stage in Table~\ref{table:veri:reward}.

We show the pseudo-code of the full mechanism in Algorithm~\ref{alg:full}. The protocol of certificate generation and verification are shown in Algorithm~\ref{alg:p2} and Algorithm~\ref{alg:v2}, respectively.

We show the incentive security properties in Section~\ref{sec:basicIS}.

\section{Discussions on Malicious Provers and Anomaly Detection}
\label{app:anomaly}

Throughout the paper, we mainly consider the scenario in which strategic provers are motivated solely by the block rewards for the training task, with their utility defined as the block reward minus computational costs. Nevertheless, in reality, there are indeed \emph{malicious} trainers who may have incentives to adversarially sabotage the model for their own benefit \cite{BytedanceAttack}. While a detailed investigation of such cases is deferred to future work, we discuss here how our mechanism could be augmented for resilience against such \emph{malicious} trainers.

\subsection{Upper Bounds on Dishonest Stages} 

To circumvent the PoL Trilemma (as discussed in Section~\ref{sec:intro}), our mechanism relaxes the requirement of Byzantine security to \emph{incentive security}. In essence, we no longer demand that the mechanism be ``absolutely secure'' against \emph{all} attacks. Instead, we only require it to be ``secure enough'' so that an attack is detected with sufficiently high probability to deter \emph{rational} players from attacking. Consequently, for an attack that is ``less severe'' and yields small utility to the attacker, even a relatively small detection probability can suffice to ensure incentive security.

A potential concern with this model is the possible \emph{underestimation} of the incentives to attack, as \emph{malicious} players may have external motivations to benefit from training an incorrect model. In that case, an attacker might still find it worthwhile to mount an attack if the benefits from corrupting the model outweigh the lost block rewards, provided that a dishonest PoL can pass verification with non-negligible probability. Nonetheless, while our security notion is relaxed, it still essentially preserves Byzantine security in most practical settings: as long as the number of dishonest stages is not too small, our mechanism can detect the attack with overwhelming probability. Formally,

\begin{proposition}
    In our full mechanism of Algorithms~\ref{alg:p2}-\ref{alg:v2}, if the prover cheats in more than $\frac{2T}{\alpha} \ln \frac{1}{\epsilon}$ stages, then the probability of passing verification (by one verifier) is at most $\epsilon$.
\end{proposition}
\label{prop:cheat:bound}

The proof of Proposition~\ref{prop:cheat:bound} is deferred to Appendix~\ref{proof:cheat:bound}.  
From the proposition, we see that our mechanism effectively preserves Byzantine security against attacks involving more than $\Theta\bigl(\frac{T}{\alpha}\bigr)$ dishonest stages. Therefore, if compromising only a small number of stages cannot substantially degrade the trained model, then any model that passes verification in our PoL mechanism can be considered effectively correct.

In particular, if we set $\alpha = \Theta(T)$ (i.e., allowing a constant-ratio overhead in the mechanism), then an adversary can only corrupt a constant number of stages with a non-negligible probability of passing the verification.

\subsection{Approaches for Anomaly Detection}

From the above discussion, we demonstrate that our mechanism effectively limits the number of dishonest stages in a PoL that can pass verification. Consequently, if we can ensure that each dishonest stage is unable to significantly corrupt the output model, we can guarantee the correctness of the trained model even in the presence of (potentially irrational) malicious provers.

Most existing work addressing this issue falls in the scope of \emph{anomaly detection}, whose primary aim is to detect significant errors at low cost \citep{paudice2018detection}. In the context of PoL, we want to ensure that the weight updates from dishonest stages do not deviate excessively from the correct updates, so that the final model remains close to one trained honestly. Although more sophisticated approaches may exist, a simple strategy is to monitor the magnitudes of weight updates: under the smoothness conditions typical of many ML problems, gradients are not expected to grow arbitrarily large. Therefore, if verifiers observe unexpectedly large updates in certain stages, they would prioritize verifying those stages to detect potential attacks (similar to \citep{jia2021proof}).

Nevertheless, in our original PoL mechanism, the verifier does not receive model weights until they select which stages to verify and obtain the corresponding weights from the prover, thus saving communication costs. To address this limitation, the PoL certificate can be augmented with a \emph{compressed} representation of the model weights that approximates the relevant distance information. According to the Johnson–Lindenstrauss lemma (Lemma~\ref{lemma:jl}), this representation can be realized via a random low-dimensional projection. The projection direction is determined by the hash of the original PoL certificate, ensuring that it cannot be manipulated or known in advance before the training is completed.

\begin{lemma}[Johnson–Lindenstrauss]
\label{lemma:jl}
    Let $X$ be a set of $n$ points in $\mathbb{R}^D$. Consider a random projection from $\mathbb{R}^D$ to $\mathbb{R}^d$ where $d = \Theta\!\bigl(\tfrac{\log n}{\epsilon^2}\bigr)$. With high probability, this projection preserves all pairwise Euclidean distances in $X$ up to a multiplicative factor of $(1 \pm O(\epsilon))$.
\end{lemma}

With this augmentation, we propose an approach to limit the effects of each dishonest stage to the output model, in order to ensure the model correctness in our PoL mechanism. We leave the detailed implementation and analysis for future work.

%\subsection{Incentive Analysis for Malicious Provers}

\section{Omitted Proofs}

\subsection{Proof of Theorem~\ref{thm:verifier:dilemma}}\label{app:proof:dilemma}

Assume we have such a mechanism. By the definition of Nash equilibrium, we consider a fixed verifier. Given that the prover and all other verifiers (if exist) act honestly, that verifier should be incentivized to do the honest verification.

Since the prover is honest, when that verifier performs honest verification, the result should always be ``Success''. However, if the verifier simply reports ``Success'' without verification, the outcome is the same but the verifier saves the verification cost, so that the verifier is incentivized to deviate from the honest strategy.

That leads to a contradiction. So no such mechanism exists.

\subsection{Proof of Theorem~\ref{thm:veri:neg}}\label{app:veri:neg}

Notice that if the verifier verifies at least one stage, then she has a computational cost of $\frac{M}{T}$.

If $v_+\le v_0$, then the verifier does not have any incentive to find a cheat, so her strict optimal strategy is reporting ``Success''. Now we assume $v_+ > v_0$.

If the verifier verifies at least one stage, then as the probability that the proof is dishonest is at most $\epsilon$, she catches a cheat with a probability upper bounded by $\epsilon$. Therefore, her expected utility is at most $v_+ \epsilon + v_0(1-\epsilon) - \frac{M}{T}$.

If the verifier just report ``Success'', her utility is $v_0$.

Since $\epsilon < \frac{M}{T(v_+-v_0)}$, we have

\begin{equation}
    v_0 > v_+ \epsilon + v_0(1-\epsilon) - \frac{M}{T}.
\end{equation}

Therefore the verifier's strict optimal strategy is to report ``Success'' without actual verification.

\subsection{Proof of Theorem~\ref{thm:main}}
\label{app:main:proof}

We assume $\alpha\ge 2$. From Eq.~\ref{eqn:utility} and $Q(1)=1$ we see that 

\begin{equation}
    u(1) = P(1)R - \int_0^1 P(x) dx \cdot M.
\end{equation}

So Eq.~\eqref{eqn:ir} implies that $u(1)>0$, i.e. the mechanism is IR. From $P(x)\ge P(1)$ we also deduce that $M<R$, i.e. the reward must be greater than the honest computation cost.

Now we estimate $Q(\rho)$ for $\rho\in [0,1)$. 

Since each cheating stage has an independent $\kappa$ probability to be caught when verified, we can equivalently model that the verification of each stage has an independent $\kappa$ probability to be \emph{effective}. In other words, a cheating stage is caught if and only if it is verified and the verification happens to be effective.

Then, we denote $\alpha^\#$ as a random variable of the total number of effectively verified stages. Hence we have:

\begin{equation}
    \Pr[\alpha^\#=s] = \binom{\alpha}{s} \kappa^s(1-\kappa)^{\alpha-s}.
\end{equation}

For the total of $T$ stages, there are $\rho T$ stages trained honestly, and $Q(\rho)$ is the probability that all $\alpha^\#$ effectively verified stages are honest. Denote $Q_s(\rho)$ as the conditional probability that the proof passes the verification given $\alpha^\#=s$, then

\begin{align}
    Q_s(\rho) &= \frac{\binom{\rho T}{s}}{\binom{T}s} \\
&= \frac{\rho T (\rho T -1) \cdots (\rho T - s +1)} { T (T -1) \cdots (T - s +1)} \\
& \le \frac{\rho T (\rho T -\rho) \cdots (\rho T - (s +1)\rho )} { T (T -1) \cdots (T - s +1)} \\
& = \rho^s.
\end{align}

Therefore, we have

\begin{align}
    Q(\rho) &= \sum_{s=0}^\alpha \Pr[\alpha^\#=s] Q_s(\rho) \\
    &\le \sum_{s=0}^\alpha \binom{\alpha}{s} \kappa^s(1-\kappa)^{\alpha-s} \rho^s \\
    &= \sum_{s=0}^\alpha \binom{\alpha}{s} (\kappa\rho)^s(1-\kappa)^{\alpha-s} \\
    &= (1-\kappa + \kappa\rho)^\alpha.
    \label{eqn:escape:bound}
\end{align}

Let $\gamma = 0$, from Eq.~\eqref{eqn:utility} we see that 

\begin{align}
u(\rho) &= P(\rho)(Q(\rho) - \gamma (1-Q(\rho)))R -   \int_0^{\rho} P(x) dx \cdot M \\
&\le (1-\kappa + \kappa\rho)^\alpha P(\rho) R - \int_0^{\rho} P(x) dx \cdot M,
\end{align}

with equality holding at $\rho = 1$.

\def\uu{\overline{u}}

Now we define $\beta = \frac{M}{R}\in (0,1)$\footnote{It is obvious that $M<R$ because the strict IR condition implies that the reward for training must be greater than the cost.} and 
\begin{equation}\label{eqn:uu}
\uu(\rho) = (1-\kappa + \kappa\rho)^\alpha P(\rho) - \beta \int_0^{\rho} P(x) dx.
\end{equation}

Notice that $P(\cdot)$ is a non-increasing function, so for $x\in [0,\rho]$, $P(x)\ge P(\rho)$. Hence, we have

\begin{align}
\uu(\rho) &= (1-\kappa + \kappa\rho)^\alpha P(\rho) - \beta \int_0^{\rho} P(x) dx \\
&\le (1-\kappa + \kappa\rho)^\alpha P(\rho) - \beta \int_0^{\rho} P(\rho) dx \\
&= ((1-\kappa + \kappa\rho)^{\alpha}-\beta \rho) P(\rho).\label{eqn:uurho}
\end{align}

Since $\rho$ is defined as the fraction of honest stages, which in practice must be multiples of $\frac{1}{T}$, we only need to prove that if Eqs.~\eqref{eqn:ir}-\eqref{eqn:bis2} hold, then

\begin{equation}\label{eqn:main:cond}
\forall \rho\in \{0\}\cup [\frac{1}{T},1), \quad \uu(\rho)<\uu(1).
\end{equation}

Now we prove \eqref{eqn:main:cond} for $\rho=0$, $\rho\in [\frac{1}{T},\frac{1}{2}]$, and $\rho\in (\frac{1}{2},1)$, respectively.

\textbf{(i) Case of $\rho=0$.}

Since $\rho=0$, we have $\uu(0) = (1-\kappa)^\alpha P(0) = (1-\kappa)^\alpha$. From \eqref{eqn:ir} we see that $\uu(0) < \uu(1)$.

\textbf{(ii) Case of $\rho\in [\frac{1}{T},\frac{1}{2}]$.}

From \eqref{eqn:uurho} we only need to prove $(1-\kappa + \kappa\rho)^{\alpha}-\beta \rho \le 0$ to deduce $\uu(\rho)\le 0 < \uu(1)$.

Define 
\begin{equation*}
    \psi(\rho) = (1-\kappa + \kappa\rho)^{\alpha}-\beta \rho.
\end{equation*} 

From $\alpha\ge 2$ we get $\psi''(\rho) = (1-\kappa + \kappa\rho)^{\alpha-2}\ge 0$, so $\phi(\cdot)$ is concave and we only need to show $\psi(\frac{1}{T})\le 0$ and $\psi(\frac{1}{2})\le 0$.

Actually, for $\rho\in\left[\frac{1}{T},\frac{1}{2}\right]$ we have

\begin{align}
\psi\left(\rho\right) &= \left(1-\kappa + \kappa\rho\right)^\alpha - \beta \rho \\
&\le \left(1 - \kappa + \frac{\kappa}{2}\right)^\alpha - \frac{\beta}{T} \\
&\le e^{-\frac{\kappa}{2}\alpha} - \frac{\beta}{T}\\
&\le e^{-\frac{\kappa}{2}\cdot \frac{2\ln \frac{T}{\beta}}{\kappa}}-\frac{\beta}{T} \\
&\le e^{-\ln \frac{T}{\beta}}-\frac{\beta}{T} \\
    &=0.
\end{align}

\textbf{(iii) Case of $\rho\in(\frac{1}{2},1)$.}

From Eq.~\eqref{eqn:uu} we get

\begin{equation}
\uu'(\rho) = \alpha\kappa (1-\kappa + \kappa\rho)^{\alpha-1}P(\rho) + (1-\kappa + \kappa\rho)^\alpha P'(\rho) -\beta P(\rho).
\end{equation}

From Eq.~\eqref{eqn:hazard:rate} we have $P'(\rho) \ge -\lambda P(\rho)$, hence

\begin{align}
\uu'(\rho) &\ge \alpha\kappa (1-\kappa + \kappa\rho)^{\alpha-1}P(\rho) - \lambda(1-\kappa + \kappa\rho)^\alpha  P(\rho) -\beta P(\rho) \\
&=((1-\kappa + \kappa\rho)^{\alpha-1} (\alpha\kappa - \lambda (1-\kappa + \kappa\rho)) -\beta ) P(\rho).\label{eqn:uudiff}
\end{align}

Now we define $t=1-\kappa+\kappa \rho$, then we have $\rho=\frac{t+(1-\kappa)}{\kappa}$ and

\begin{equation}
    t\in(1-\frac{\kappa}{2},1).\label{eqn:t:range}
\end{equation}

We denote

\begin{align*}
    V(t) &= (1-\kappa + \kappa\rho)^{\alpha-1} (\alpha\kappa - \lambda (1-\kappa + \kappa\rho)) -\beta \\
    &= -\lambda t^\alpha + \alpha \kappa t^{\alpha-1} - \beta, \\
    U(t) &= ((1-\kappa + \kappa\rho)^{\alpha}-\beta \rho) \\
    &= t^\alpha - \frac{\beta}{\kappa} t + \frac{\beta(1-\kappa)}{\kappa},
\end{align*}

then from \eqref{eqn:uudiff} we see that

\begin{equation}
    \uu'(\rho) \ge V(t)P(\rho), \label{eqn:uudiff2}
\end{equation}

and from \eqref{eqn:uurho} we see that

\begin{equation}
    \uu(\rho) \le U(t) P(\rho).\label{eqn:uu:ub}
\end{equation}

\def\uuu{\overline{\uu}}

For $\rho\in [\frac{1}{2},1]$ we define 
\[
    \uuu(\rho) = \uu(1) - \int_{\rho}^1 V(1-\kappa+\kappa x)P(x) dx,
\]

then

\begin{equation}
    \uuu'(\rho) = V(t)P(\rho) \le \uu'(\rho). \label{eqn:uuudiff}
\end{equation}

and from \eqref{eqn:uudiff2} we deduce

\begin{align}
    \uuu(\rho) &= \uu(1)- \int_{\rho}^1 \uuu'(x) dx \\
    &\ge \uu(1) - \int_{\rho}^1 \uu'(x) dx \\
    &=\uu(\rho).
\end{align}

From \eqref{eqn:bis2} we have $\alpha \ge \frac{2 (\lambda+\beta)}{\beta\kappa} \ge \frac{\lambda}{\kappa}+1$, thus we get 

\begin{align}
V'(t)&=\alpha t^{\alpha-2}((\alpha-1)\kappa-\lambda t) \\
&\ge \alpha t^{\alpha-2} (\lambda-\lambda t) \\
&\ge 0.
\end{align}

Hence $V(t)$ has at most one zero point on $(\frac{1}{2},1)$, and from \eqref{eqn:uuudiff}, $\uuu(\rho)$ has at most one stationary point on $(\frac{1}{2},1)$. Because $V(1)=-\lambda+\alpha\kappa-\beta\ge 0$, we deduce that $\uuu(\rho)$ must satisfy one of the following:

\begin{itemize}
\item Monotonic increasing on $(\frac{1}{2},1)$, or
\item Monotonic decreasing on $(\frac{1}{2},\xi)$ and increasing on $(\xi,1)$, in which $\xi \in (\frac{1}{2},1)$.
\end{itemize}

In the first case, it holds that $\uu(\rho)\le \uuu(\rho) < \uuu(1) = \uu(1)$ for $\rho\in(\frac{1}{2},1)$ and we prove \eqref{eqn:main:cond}. Now we consider the second case.

Since $\uuu(\rho)$ is increasing on $(\xi,1)$, we see that $\forall \rho\in[\xi,1)$, $\uu(\rho)\le\uuu(\rho) < \uuu(1)=\uu(1)$. On the other hand, when $\rho\in(\frac{1}{2},\xi)$, we prove that $\uu(\rho)\le 0$.

Actually, because $\uuu(\cdot)$ is decreasing at $\rho \in(\frac{1}{2},\xi)$, we deduce that $\uuu'(\rho) \le 0$, thus from \eqref{eqn:uuudiff} we have $V(t)\le 0$. 

Additionally, we have
\begin{small}
\begin{align}
    &t V(t)-\alpha\kappa U(t) \\
    &= (-\lambda t^{\alpha+1} + \alpha\kappa t^\alpha -\beta t) - (\alpha \kappa t^\alpha -\alpha\beta t + \alpha \beta (1-\kappa)) \\
    &= -\lambda t^{\alpha+1} -\beta t + \alpha\beta t - \alpha\beta (1-\kappa) \\
    &= \alpha \beta (t+\kappa - 1) - (\lambda t^{\alpha+1} +\beta t).
\end{align}
\end{small}

From \eqref{eqn:bis2} we have $\alpha \ge \frac{2 (\lambda+\beta)}{\beta\kappa}$, and from \eqref{eqn:t:range} we have $1-\frac{\kappa}{2} < t < 1$. Therefore,

\begin{align}
    t V(t)-\alpha\kappa U(t) &> \frac{2(\lambda+\beta)}{\beta\kappa} \beta \left(1-\frac{\kappa}{2}+\kappa-1\right) - (\lambda+\beta) \\
    &= \frac{2(\lambda+\beta)}{\kappa}\cdot \frac{\kappa}{2} - (\lambda+\beta) \\
    &= 0.
\end{align}

Combined with $V(t)\le 0$, we deduce that $U(t)\le 0$, and from \eqref{eqn:uu:ub} we get $\uu(\rho)\le 0 < \uu(1)$.

Here we finish the proof for all three cases of \eqref{eqn:main:cond}. Now we have proven Theorem~\ref{thm:main}.

\subsection{Proof of Theorem~\ref{thm:penalty}}
\label{app:proof:penalty}

It is straightforward to see that Eq.~\eqref{eqn:ir2} holds if and only iff the mechanism is IR. Similar to the proof in Appendix~\ref{app:main:proof}, we have 

\begin{align}
u(\rho) &= P(\rho)(Q(\rho) - \gamma (1-Q(\rho)))R -   \int_0^{\rho} P(x) dx \cdot M \\
&\le ((1+\gamma)(1-\kappa + \kappa\rho)^\alpha - \gamma) P(\rho) R - \int_0^{\rho} P(x) dx \cdot M.
\end{align}

Hence, we can similarly define

\begin{equation}\label{eqn:uu2}
\uu(\rho) = ((1+\gamma)(1-\kappa + \kappa\rho)^\alpha-\gamma) P(\rho) - \beta \int_0^{\rho} P(x) dx,
\end{equation}

and only need to prove that

\[
\uu(\rho)<\uu(1), \quad \rho\in [0,1).
\]

For Eq.~\eqref{eqn:uu2} we see that

\begin{equation}\label{eqn:penalty:division}
    (1-\kappa+\kappa\rho)^\alpha \le \frac{\gamma}{1+\gamma } \implies \uu(\rho) \le 0.
\end{equation}

Now we consider two cases of $(1-\kappa)^\alpha < \frac{\gamma}{1+\gamma}$ and $(1-\kappa)^\alpha \ge \frac{\gamma}{1+\gamma}$ separately.

\textbf{(i) Case of $(1-\kappa)^\alpha < \frac{\gamma}{1+\gamma}$.}

\def\rhoth{\rho_{th}}

In this case, we define $\rhoth = \frac{(\frac{\gamma}{1+\gamma})^{\frac{1}{\alpha}}+\kappa-1}{\kappa}$, then for $\rho\in [0,1]$, we have 

\begin{equation}\label{eqn:rhoth}
\rho \le \rhoth \iff (1-\kappa+\kappa\rho)^\alpha \le \frac{\gamma}{1+\gamma}.
\end{equation}

From Eq.\eqref{eqn:penalty:division} and IR guarantee we have that $\uu(\rho)\le 0 < \uu(1)$ when $\rho \in [0,\rhoth]$. Now we consider $\rho \in (\rhoth,1).$

From Eq.\eqref{eqn:uu2} we have 
\begin{small}
\begin{align}
\uu'(\rho) &= \alpha\kappa (1+\gamma) (1-\kappa + \kappa\rho)^{\alpha-1}P(\rho)\nonumber\\
&\qquad + ((1+\gamma)(1-\kappa + \kappa\rho)^\alpha -\gamma)P'(\rho) -\beta P(\rho) \\
&\ge ((1+\gamma)(\alpha\kappa (1-\kappa + \kappa\rho)^{\alpha-1} \nonumber\\
&\qquad - \lambda (1-\kappa + \kappa\rho)^\alpha )+\lambda\gamma-\beta) P(\rho) \\
&= ((1+\gamma)(1-\kappa+\kappa\rho)^{\alpha-1}(\alpha\kappa-\lambda(1-\kappa+\kappa\rho))+\lambda\gamma-\beta) P(\rho).
\end{align}
\end{small}
From \eqref{eqn:bis3} we have $\alpha\kappa \ge \lambda$, hence

\begin{equation}
    \alpha \kappa - \lambda(1-\kappa+\kappa\rho) \ge 0.
\end{equation}

From Eq.~\eqref{eqn:rhoth} and $1-\kappa+\kappa\rho \in [0,1]$, we have

\begin{align*}
\rho > \rhoth &\implies (1-\kappa+\kappa\rho)^\alpha \ge\frac{\gamma}{1+\gamma} \\
&\implies (1-\kappa+\kappa\rho)^{\alpha-1} \ge\frac{\gamma}{1+\gamma}.
\end{align*}

Therefore, for $\rho\in (\rhoth, 1)$, we have

\begin{align}
\uu'(\rho) &\ge ((1+\gamma)\cdot \frac{\gamma}{1+\gamma}\cdot(\alpha\kappa-\lambda(1-\kappa+\kappa\rho)) + \lambda\gamma-\beta)P(\rho) \\
&=(\gamma (\alpha\kappa-\lambda(1-\kappa+\kappa\rho)) +\lambda\gamma-\beta)P(\rho) \\
&\ge(\gamma(\alpha\kappa - \lambda)+\lambda\gamma -\beta)P(\rho) \\
&=(\alpha\gamma\kappa-\beta)P(\rho).
\end{align}

From \ref{eqn:bis3} we have $\alpha > \frac{\beta}{\gamma\kappa}$, and as $\gamma,\kappa>0$, we have $\alpha\gamma\kappa-\beta$>0, hence $\uu'(\rho)>0$. 

Therefore, $\uu(\cdot)$ is monotonic increasing on $(\rhoth,1)$, deducing that $\uu(\rho)<\uu(1)$ for $\rho\in (\rhoth,1)$. 

\textbf{(ii) Case of $(1-\kappa)^\alpha \ge \frac{\gamma}{1+\gamma}$}.

In this case, we have $(1-\kappa+\kappa\rho)^\alpha\ge \frac{\gamma}{1+\gamma}$ for $\rho \in [0,1)$, so it holds that $\uu(\cdot)$ is monotonic increasing on $[0,1)$. Hence, we prove that $\uu(\rho)<\uu(1)$ for $\rho\in [0,1)$. 

\subsection{Proof of Lemma~\ref{lem:verifier:util}}
\label{app:verifier:util}

Because it is a $(0,n-m)$-collusion in which the prover is honest, and the number of dishonest verifiers does not exceed $n-m<m$, a report will be \acc ed if and only if it is correct. 

We consider the three cases in which the verifier is honest, lazy, and malicious, respectively.

\textbf{Honest verifier.} When the verifier is honest, as we assume the prover is honest, the stage has a flag with probability $\eta$, with a probability $\frac{\eta}{2}$ of $F_1$ and $F_2$ each.

If there is no flag, then the verifier gets a reward of $0$ while paying a verification cost of $\frac{M}{T}$; if there is a flag, she gets a reward of $R_1$ while paying $\frac{2M}{T}$ as she needs to check two seeds (as described in Section~\ref{subsec:CTF}.) Hence, the expected utility of an honest verifier is
\begin{equation}
    (1-\eta)\cdot (-\frac{M}{T}) + \eta\cdot (R_1-2\frac{M}{T})=\eta R_1-(1+\frac{\eta}{2})\frac{M}{T}.
\end{equation}

\textbf{Lazy verifier.} If she is lazy, then she may report ``Normal'', $F_1$, $F_2$, or ``Invalid''. 

If she reports ``Normal'', she gets a $0$ reward when there is not a flag, and $-2R_1$ when there is a flag, so the expected utility is
\begin{equation}
    \eta \cdot (-2R_1) = -2\eta R_1.
\end{equation}

If she reports $F_1$ or $F_2$, then she gets $R_1$ with probability of $\frac{\eta}{2}$ and gets $-2R_1$ with probability of $1-\frac{\eta}{2}$. Hence, the expected utility is  
\begin{equation}
    \frac{\eta}{2}R_1+(1-\frac{\eta}{2}) \cdot (-2R_1) = -(2-\frac{\eta}{2}) R_1.
\end{equation}

If the verifier reports ``Invalid'', since the prover is honest, the verifier gets a reward of $-2R_1\le \max\{-2\eta R_1,-(2-\frac{\eta}{2}) R_1\}$.

Hence, a lazy verifier can get an expected utility no more than $\max\{-2\eta R_1,-(2-\frac{\eta}{2}) R_1\}$.

\textbf{Malicious verifier.} If the verifier is malicious, then she has verified but deliberately reports an incorrect answer. Then, similar to the first case, she pays an expected verification cost of $(1+\frac{\eta}{2})\frac{M}{T}$ and receives a reward of $-2R_1$. The overall utility is $-2R_1-(1+\frac{\eta}{2})\frac{M}{T} \le -2R_1 \le \max\{-2\eta R_1,-(2-\frac{\eta}{2}) R_1\}.$

\subsection{Proof of Theorem~\ref{thm:vis}}
\label{app:proof:vis}

%\added{[REWORK!]}

Since $\eta<\frac{4}{5}$, from Table~\ref{table:veri:util}, the verifier's utility of being dishonest when the prover is honest is at most $\max\{-2\eta R_1,-(2-\frac{\eta}{2})R_1\}=-2\eta R_1$. Hence, when an $\eps$ fraction of provers are dishonest, even when there are at most $n-m$ dishonest verifiers, the expected utility of a verifier being honest is lower bounded by:
\begin{equation}
    u_h=(1-\eps) \cdot(\eta-(1+\frac{\eta}{2}c)) R_1 + \eps\cdot (-2R_1),
\end{equation}
and the utility of being dishonest is lower bounded by:
\begin{equation}
    u_d=(1-\eps) \cdot(-2\eta R_1) + \eps\cdot R_1.
\end{equation}

To make the mechanism $\eps$-VIS, we only need $u_h>0$ and $u_d<0$, hence we only need 
\begin{equation}
    \eta < \min\left\{\frac{\eta-(1+\frac{\eta}{2})c}{\eta-(1+\frac{\eta}{2})c+2},\frac{2\eta}{2\eta+1}\right\}.
\end{equation}

Notice that $\eta<1$, hence 
\begin{align}
    \min\left\{\frac{\eta-(1+\frac{\eta}{2})c}{\eta-(1+\frac{\eta}{2})c+2},\frac{2\eta}{2\eta+1}\right\} &> \min\left\{\frac{\eta-(1+\frac{\eta}{2})c}{3},\frac{2\eta}{3}\right\} \\
    &\ge \frac{\eta-(1+\frac{\eta}{2})c}{3}.
\end{align}

Therefore, as long as $\eta <\frac{4}{5}$ and $\eps \le \frac{\eta-(1+\frac{\eta}{2})c}{3}$, the mechanism is $\eps$-VIS even in the presence of at most $n-m$ dishonest verifiers.

Since we desire $\eta<\frac{4}{5}$ and $\frac{\eta-(1+\frac{\eta}{2})c}{3}>0$, a $c<\frac{4}{7}$ ensures the existence of such $\eta$.

\subsection{Proof of Proposition~\ref{prop:cheat:bound}}
\label{proof:cheat:bound}

From Eq.~\eqref{eqn:escape:bound} in Appendix~\ref{app:main:proof}, denoting $\rho$ as the fraction of honestly trained stages, the probability of passing the verification is
\begin{equation}
    Q(\rho) \le (1-\kappa+\kappa\rho)^\alpha.
\end{equation}

In our full mechanism we have $\kappa=\frac{1}{2}$, and denote $\Delta$ as the number of dishonest stages, then we have $\rho = 1-\frac{\Delta}{T}$. Hence, we deduce that
\begin{align}
    Q(\rho)&\le \left(1-\frac{\Delta}{2T}\right)^\alpha\\
    &\le e^{-\frac{\alpha}{2T}\cdot \Delta}.
\end{align}

Since $\Delta \ge \frac{2T}{\alpha}\ln\frac{1}{\epsilon}$, we have
\begin{align}
    Q(\rho)
    &\le e^{-\frac{\alpha}{2T}\cdot \frac{2T}{\alpha}\ln\frac{1}{\epsilon}}\\
    &= e^{-\ln \frac{1}{\epsilon}}\\
    &=\epsilon.
\end{align}

%\revise{[TO BE ADDED. (1) is true but a little tricky to rigorously prove.]}

%\section{Additional Experimental Results}

%\subsection{Verifier's Incentive in Constant Strategy}
%\subsection{Figures Omitted in Main Text}
%\label{app:vi:cs}

%\subsection{Distribution of $\alpha'$ in Greedy-Adaptive Strategy}

%\clearpage

\end{document}
\endinput
%%
%% End of file `sample-sigconf.tex'.